\documentclass[useAMS,usenatbib]{mn2e}
\usepackage{graphicx}
\usepackage{epsfig}
\usepackage{amsmath}
\usepackage{mathrsfs}
\usepackage{amssymb}
\usepackage[english]{babel}
\usepackage{amsfonts}
\usepackage{fontenc}
\usepackage{times}
\usepackage{txfonts}
\usepackage{textcomp}
\usepackage{booktabs}
\usepackage{microtype}
\usepackage{aas_macros}
\usepackage{xspace}
\usepackage[utf8]{inputenc}
\usepackage[T1]{fontenc}
\usepackage[pdftex]{lscape}
\usepackage[usenames,dvipsnames,svgnames,table]{xcolor}
\usepackage{flushend}

\newcommand*{\se}{\sigma_{\rm e}}
\newcommand*{\re}{R_{\rm e}}
\newcommand*{\Ae}{A_{\rm e}}
\newcommand*{\remaj}{R_{\rm e}^{\rm maj}}
\newcommand*{\fDM}{f_{\rm DM}}
\newcommand*{\Mein}{M_{\rm EIN}}
\newcommand*{\Rein}{R_{\rm EIN}}
\newcommand*{\mI}{m_{I}}
\newcommand*{\Mr}{M_{r}}
\newcommand*{\MV}{M_{V}}
\newcommand*{\Msun}{M_{\odot}}
\newcommand*{\Lsunr}{L_{\odot\,r}}
\newcommand*{\MLdyn}{(M_*/L)_{\rm dyn}}
\newcommand*{\MLpop}{(M_*/L)_{\rm Salp}}
\newcommand*{\MLself}{(M/L)_{\rm MFL}}
\newcommand*{\kms}{\mathrm{km~s}^{-1}}
\newcommand*{\Mjam}{M_{\rm JAM}}
\newcommand*{\ATLAS}{ATLAS$^{\mathrm{3D}}$}

\title[The stellar initial mass function of early-type galaxies]
{The stellar initial mass function of early type galaxies from low to high stellar velocity dispersion:
homogeneous analysis of \ATLAS and Sloan Lens ACS galaxies}
 \author[S. Posacki et al.]{Silvia Posacki$^{1}$\thanks{E-mail:
silvia.posacki@unibo.it}, Michele Cappellari$^{2}$, Tommaso Treu$^{3,4}$, Silvia Pellegrini$^{1}$ \& Luca Ciotti$^{1}$
\\$^{1}$Department of Physics and Astronomy, University of Bologna, viale Berti Pichat 6/2, 40127 Bologna, Italy\\
$^{2}$Sub-department of Astrophysics, Department of Physics, University of Oxford, Denys Wilkinson Building, Keble Road, Oxford OX1 3RH\\
$^{3}$Department of Physics, University of California, Santa Barbara, CA 93106, USA\\
$^{4}$Department of Physics and Astronomy, University of California, Los Angeles, CA 90095, USA}

\date{Accepted 2014 October 7}

\pagerange{\pageref{firstpage}--\pageref{lastpage}} \pubyear{2014}

\begin{document}
\maketitle
\label{firstpage}

\begin{abstract}
We present an investigation about the shape of the initial mass function (IMF) of early-type galaxies (ETGs), based on a joint lensing and dynamical analysis, and on stellar population synthesis models, for a sample of 55 lens ETGs identified by the Sloan Lens ACS (SLACS) Survey. 
We construct axisymmetric dynamical models based on the Jeans equations which allow for orbital anisotropy and include a
dark matter halo. The models reproduce in detail the observed \textit{HST} photometry and are constrained by the total
projected mass within the Einstein radius and the stellar velocity dispersion ($\sigma$) within the SDSS fibers.
Comparing the dynamically-derived stellar mass-to-light ratios $\MLdyn$, obtained for an assumed halo slope
$\rho_{\rm h}\propto r^{-1}$, to the stellar population ones $\MLpop$, derived from full-spectrum fitting and assuming
a Salpeter IMF, we infer the mass normalization of the IMF.
Our results confirm the previous analysis by the SLACS team that the mass normalization of the IMF of high $\sigma$ galaxies is consistent on average with a Salpeter slope.
Our study allows for a fully consistent study of the trend between IMF and $\sigma$ for both the SLACS and \ATLAS samples, which explore quite different $\sigma$ ranges.
The two samples are highly complementary, the first being essentially $\sigma$ selected, and the latter volume-limited and nearly mass selected.
We find that the two samples merge smoothly into a single trend of the form 
$\log\alpha =(0.38\pm0.04)\times\log(\se/200\,\kms)+(-0.06\pm0.01)$, where $\alpha=\MLdyn/\MLpop$ and $\se$ is the luminosity averaged $\sigma$ within one effective radius $\re$.
This is consistent with a systematic variation of the IMF normalization from Kroupa to Salpeter in the interval $\se\approx90-270\,\kms$.

\end{abstract}


\section{Introduction} \label{sec:intro}
The stellar initial mass function (IMF) describes the mass distribution of the stellar population originated in a single star formation burst, at the time of birth.
It gives us information about the relative importance of low and high mass stars, hence its form directly affects the amount of stellar ejecta and their chemical composition, the mass distribution of stellar remnants, and the stellar mass-to-light ratio of the population. The study of the shape of the IMF also gives us direct insights into the physics of star formation, and it is crucial for the estimate of galaxy stellar masses starting from the observed luminosity.
Thus the knowledge of the IMF is fundamental in many fields of astrophysics that study the formation and evolution of stellar systems.
Several direct measurements of star counts of resolved stellar populations in the solar neighbourhood have shown that the IMF can be parametrized by a power law mass distribution $dN/dM\propto M^{-s}$, characterized by a \citet{Salpeter.1955} slope $s\simeq2.35$ for $M\gtrsim0.5\Msun$, and by a change toward flatter slopes for $M\lesssim0.5\Msun$ \citep{Kroupa.2001,Chabrier.2003}. This holds in different environments throughout the Milky Way \citep{Kroupa.2002,Bastian.etal2010}, but whether this is true for all galaxies is still ongoing debate.
Stellar counts down to very low stellar masses (i.e., in the mass range of major uncertainty given the intrinsic difficulty of measurements) is not feasible in distant external galaxies, so that, in order to study the extragalactic IMF, people use alternative methods based, for example, on ionized gas emission, redshift evolution of the tilt and normalization of the Fundamental Plane, strength of IMF-sensitive spectral features, gas kinematics, gravitational lensing and stellar dynamics (see \citealt{Cappellari.etal2013XX} for a more detailed review). 
Among these indirect methods, it is widely used to constrain the IMF shape by estimating galaxy stellar masses from dynamical models and comparing them with the predictions of stellar population synthesis models, that rely on an assumed IMF shape.
Note that this method does not directly measure the shape of the IMF, but its overall mass normalization: each IMF shape results in a different $M_*/L$, that is converted in a different stellar mass, once the luminosity is measured.
In the last decade a number of works based on this method have agreed that spiral galaxies are inconsistent with a Salpeter normalization over the whole mass range, and that they need a lighter overall normalization similar to Kroupa or Chabrier, like the Milky Way \citep{Bell.deJong2001,Kassin.etal2006,Bershady.etal2011,Brewer.etal2012}. The same result also appears to be valid for at least some ETGs \citep{Cappellari.etal2006,Ferreras.etl2008,Dutton.etal2011,Thomas.etal2011,Brewer.etal2014,Zepf.etal2014}, thus showing no evidence of a departure from a universal stellar IMF.

In contrast, however, there are numerous works carried out on ETG samples that point out evidences of a dynamical mass excess over the predictions of stellar population models with fixed IMF. This excess increases with galaxy mass and it can be explained either (i) by an IMF normalization that increases from a Kroupa/Chabrier one at low masses, up to a Salpeter normalization for the more massive galaxies, implying a systematic variation of the IMF (e.g., \citealp{Renzini.Ciotti.1993}), or (ii) by an increase of the dark matter (DM) fraction as function of galaxy mass due to a non-universal DM halo profile (\citealp[hereafter T10]{Padmanabhan.etal2004,Cappellari.etal2006,Grillo.etal2009,Thomas.etal2009,Tortora.etal2009,Auger.etal2010,Graves.Faber2010,Schulz.etal2010,Treu.etal2010}; \citealp{Barnabe.etal2011,Dutton.etal2012,Tortora.etal2013}).
This method, based on the comparison between galaxy stellar masses computed from dynamical and stellar population synthesis models, is indeed subject to degeneracies in the dynamical modelling, which are related to the assumptions for the luminous and dark matter density profiles, and for the velocity dispersion anisotropy. However, the degeneracies can be reduced by additional constraints derived, for example, from gravitational lensing analysis or integral field spectroscopy observations.

An example is given by the results of the SLACS group: T10 analysed 56 lens ETGs belonging to the SLACS sample by building dynamical models tuned to reproduce the SDSS-measured velocity dispersion $\sigma_*$ and the total projected mass within the Einstein radius. They adopted two-component spherical isotropic dynamical models with self-similar \citet{Hernquist.1990} profiles to describe the stellar density, and a NFW \citep{Navarro.etal1997} DM density distribution with fixed slope and scale radius. T10 found that bottom-heavy IMFs such as Salpeter are strongly preferred over light-weight IMFs such as Kroupa/Chabrier for the most massive ETGs, assuming standard NFW dark matter density profiles. 
This result was then strengthened by \citet{Auger.etal2010} who included adiabatic contraction and weak-lensing constraints in the modelling, and found that only Salpeter-like IMF are consistent with the observed properties of their ETG sample. 
Note this is an effectively velocity dispersion selected sample, so that it is composed of high $\sigma$ galaxies (see Section~\ref{sec:data} and references therein).

Another remarkable example is the work of \citet{Cappellari.etal2012,Cappellari.etal2013XV,Cappellari.etal2013XX} on the volume-limited, nearly mass selected \ATLAS sample of 260 ETGs. They constructed detailed axisymmetric dynamical models, which allow for orbital anisotropy and reproduce in detail both the galaxy images and the high-quality integral-field stellar kinematics out to about one effective radius $\re$. Given the tighter constraints with respect to previous analogous studies, their models were well-suited to explore different DM density profiles, and they find that a non-universal IMF is always required under all halo assumptions, due to the low DM mass contribution within $\re$.
Their study, based on an unprecedented large sample of ETGs spanning a wide range in galaxy mass, found a systematic trend in IMF normalization varying from Kroupa/Chabrier up to Salpeter or heavier for increasing velocity dispersion.

Finally other works, based on IMF-sensitive spectral features, that are completely independent of dynamical modelling assumptions, find a steepening IMF with increasing velocity dispersion and $[$Mg$/$Fe$]$, with massive ETGs requiring bottom-heavy, dwarf-rich IMF \citep{vanDokkum.Conroy2010,vanDokkum.Conroy2011,Conroy.vanDokkum2012,Spiniello.etal2012,Spiniello.etal2014,Ferreras.etal2013}.
Thus, there seems to be a systematic dependence of the IMF on galaxy properties, indicating that high mass ETGs prefer on average a Salpeter normalization, while low mass galaxies are consistent with a lighter normalization, similar to Kroupa or Chabrier.
However, quantitative consistency between the dynamical and spectral synthesis approach has not been achieved yet (e.g. \citealt{Smith.2014}; McDermid et al, submitted).

In this work we revisit the analysis of T10 in order to investigate the effects of a more detailed modelling of the stellar component. T10 spherical models indeed provide only a crude approximation to the observed galaxy surface brightness, which shows evidence for disks and it is known to vary systematically with galaxy mass \citep{Caon.etal1993}.
To address this potential bias, here we construct models which allow for axisymmetry and can reproduce the observed galaxy surface brightness in detail, in an essentially non-parametric way. Moreover, differently from T10, our stellar population synthesis models are built via full spectrum fitting of SDSS spectra, and not by means of multicolour photometry.
An approach closely related to the one illustrated in this work, was employed also by \citet{Barnabe.etal2013}
in their analysis of two SLACS ETGs, where they also exploited X-Shooter spatially-resolved kinematic data in order to
put constraints on these systems' IMFs.

Finally, our analysis is similar to that performed by \citet{Cappellari.etal2013XV}, therefore this allows us also to combine the SLACS and the \ATLAS samples, obtaining a larger and homogeneously analysed sample of ETGs. Remarkably, due to their selection criteria the two samples are complementary, so that the combined sample is fairly representative of ETGs, extending from low to very high velocity dispersions (or stellar masses). 
Another attempt to compare similar previous works was made by \citet{Dutton.etal2013}, even though it is not as homogeneous as this.

We proceed as follows. In Section~\ref{sec:data} we briefly summarize the sample and data, while in Section~\ref{sec:methods} we describe our dynamical and stellar population synthesis models. The main results are presented in Section~\ref{sec:results}, and Section~\ref{sec:conclusions} summarizes the conclusions.
All magnitudes are in the AB photometric system, and a standard concordance cosmology is assumed, i.e. $h=0.7$, $\Omega_{m}=0.3$ and $\Omega_{\Lambda}=0.7$.

\section{Sample and data} \label{sec:data}
The subsample of galaxies analysed in this work is extracted from the SLACS sample studied in T10. The SLACS sample is composed of massive ETGs, that were spectroscopically selected from the Sloan Digital Sky Survey (SDSS) database for being gravitational lenses \citep{Bolton.etal2006}.
In particular, the SLACS sample consists of galaxies with very high $\sigma$ for two main reasons: (i) the lensing cross section scales approximately with $\sigma^4$, and (ii) the SDSS is a flux-limited sample, so that high-luminosity, and therefore high $\sigma$, galaxies are overrepresented because they are visible over a larger volume \citep{Hyde.Bernardi.2009}.
Thus, the SLACS sample is effectively $\sigma$-selected \citep{Auger.etal2010,Ruff.etal2011}.
Several studies have shown that the SLACS sample is indistinguishable from a $\sigma$-selected sample of non-lens ETGs \citep{Bolton.etal2006,Treu.etal2006,Treu.etal2009}.

We selected our SLACS subsample by requiring the availability of \textit{HST} photometry in the \textit{I}-band, since it is expected to better trace the luminous mass, being less affected by the presence of dust.
In this way, we obtained a subsample of 55 galaxies that span a redshift range of $0.06\lesssim z \lesssim 0.36$.
Our data consist of \textit{HST}/ACS/\textit{F}814\textit{W} images \citep{Auger.etal2009}, and SDSS optical spectra taken from data release ten (DR10, \citealt{Ahn.etal2014}).
SDSS spectra cover the wavelength range $3800-9200$ \AA, with a spectral resolution of $\sim 2.76$ \AA~FWHM, which corresponds to an intrinsic dispersion $\sigma_{\mathrm{int}}\sim 85\,\kms$ at 3800~\AA~and $\sigma_{\mathrm{int}}\sim 50\,\kms$ at 9000~\AA.

\section{Methods} \label{sec:methods}
In order to study the mass normalization of the IMF for our sample of 55 ETGs, we compare the stellar mass-to-light ratios $M_*/L$ determined from two different and independent diagnostics of galaxy stellar mass. The first method relies on gravitational lensing and stellar kinematics, it involves the construction of dynamical models, and so it is sensitive to galaxy mass structure and stellar dynamics assumptions (Sect.~\ref{sec:dynmodel}). 
Here we try to reduce the unavoidable degeneracies generating from the assumption of a particular stellar profile, by using a parametrization which allows for a large number of free parameters.
This approach is able to reproduce the galaxy surface brightness images in detail, adding new parameters until the difference between the model and the image becomes negligible.
The second approach is instead based on stellar population synthesis models, it assumes an IMF, and returns an estimate of $M_*/L$ by means of spectral fitting; the reliability of this method depends mostly on the goodness of the stellar templates (Sect.~\ref{sec:spsmodel}).

\subsection{The dynamical modelling} \label{sec:dynmodel}

\subsubsection{The mass structure}\label{sec:mass_struc}
The mass structure of our galaxy models consists of three components: an axisymmetric stellar distribution, a spherical DM halo, and a central supermassive black hole (BH).

The stellar component is accurately modelled with the aid of \textit{I}-band \textit{HST} images on which we performed a multi gaussian expansion (MGE) axisymmetric parametrization (\citealp{Emsellem.etal1994}, see also \citealp{Bendinelli.1991,Bendinelli.etal1993}) that fits the galaxy surface brightness distribution. In particular, we used the \textsc{mge$\_$fit$\_$sectors} software package of \citet{Cappellari.2002}\footnote{Available at http://purl.org/cappellari/software}, where the MGE formalism and the fitting algorithm are fully described.
Given the nature of our sample, the galaxy images are characterized by the presence of several gravitational arches or rings that we properly mask in order to obtain a better fit.
We impose the surface brightness profile of the MGE model to decrease as $R^{-4}$ at large radii, so as to limit the inclusion of spurious light from nearby galaxies.
All the model gaussians are convolved with a gaussian point spread function with a dispersion of 0.04 arcsec, as befitting for ACS.
We also use some prescriptions for the gaussians' axial ratio: in the limits of obtaining a good fit of the surface brightness, we force 1) the flattest gaussian to have the highest axial ratio, and 2) the gaussians' axial ratio range to be the smallest possible. In this way, we both avoid an artificial restriction of the range of the possible inclination angles for which the model can be deprojected (see \citealt{Cappellari.2002}, Section 2.2.2), and we 'regularize' the model, preventing significant variations of the axial ratio, as physically plausible.
These assumptions are needed because the deprojection, to obtain the intrinsic stellar luminosity density from the observed surface brightness, is mathematically non-unique \citep{Rybicki.1987, Gerhard.Binney.1996}.
Moreover, some of the galaxies have been observed only once so that we remove the presence of cosmic rays using the \textsc{la$\_$cosmic} software of \citet{vanDokkum.2001}\footnote{Available at http://www.astro.yale.edu/dokkum/lacosmic/}. 
The model flux is corrected for foreground galactic extinction following \citet{Schlafly.etal2011}, as given by the NASA/IPAC Extragalactic Database (NED), and the apparent $I$-band magnitude $\mI$ is computed, assuming $M_{\odot\,I}=4.57$ \footnote{Taken from http://mips.as.arizona.edu/{\textasciitilde}cnaw/sun.html}. Then, by means of SDSS spectra, we perform a k-correction following \citet{Hogg.etal2002}, and we transform all observed magnitudes to consistent $V$ and $r$-band rest-frame magnitudes, $\MV$ and $\Mr$ respectively, assuming
the redshift values reported in Table~\ref{tab:mge}. This correction was necessary due to the non-negligible redshift range spanned by the galaxies, and the choice of the photometric bands is motivated by the possibility to compare our results with the SLACS and \ATLAS ones, that have been obtained in these bands.
Then, assuming $M_{\odot\,r}=4.64$ \citep{Blanton.etal2007}, we normalized the model gaussians in units of $L_{\odot\,r}\,$pc$^{-2}$, so that now the MGE model has the right units and format to be used for the dynamical modelling (see Section~\ref{sec:jam}).
The MGE models for all the 55 galaxies are shown in Fig.~\ref{fig:MGE}, Table~\ref{tab:mge} reports their
magnitudes, and all their parameters are listed in Appendix~\ref{app:mge}.
As a sanity check, we compared our $\mI$ with the ones reported by \citet{Bolton.etal2008}: their magnitudes were calculated, starting from the same data, by fitting two-dimensional ellipsoidal \citet{deVaucouleurs.1948} luminosity profiles, and are the result of the full (not truncated) analytic integral of the best fitting de Vaucouleurs model. We find that the two sets of magnitudes agree with an rms scatter of 0.08 mag, but our $\mI$ are systematically higher by 0.18 mag, implying fluxes underestimated by 18 per cent.
This is likely due to the fact that SLACS magnitudes are extrapolated to infinite radii, while ours are limited to the observed photons.
Finally, in Fig.~\ref{fig:MV_compare} we compared our $\MV$ with the magnitudes calculated by \citet{Auger.etal2009} in the same band, and found they are consistent with an rms scatter of 0.07 mag, which implies an error of 5 per cent in the luminosity; we assume the same error also for $\Mr$.

\begin{table*}
\caption{Properties and JAM models parameters of the 55 galaxy SLACS subsample.}
\begin{tabular}{cccccccccc}
\toprule   
Name        & z      &$\sigma_{*}$  &$\Rein$&$\log\Mein$  &$\mI$    & $\MV$  & $\Mr$  &$\log\remaj$ &$\log\re$   \\  
            &        &[km s$^{-1}$] & [kpc] &[$M_{\odot}$]& [mag]   & [mag]  & [mag]  &[arcsec]     &[arcsec]    \\  
(1)         & (2)    &    (3)       &  (4)  &  (5)        & (6)     & (7)    & (8)    &  (9)        &  (10)      \\  
 \midrule                                                                                                            
J0029--0055 & 0.2270 & 229 $\pm$ 18 & 3.48  & 11.08       & 17.16   & -22.67 & -22.79 & 2.017       & 1.830       \\     
J0037--0942 & 0.1955 & 279 $\pm$ 14 & 4.95  & 11.47       & 16.35   & -23.09 & -23.20 & 2.016       & 1.763       \\     
J0044+0113  & 0.1196 & 266 $\pm$ 13 & 1.72  & 10.96       & 15.83   & -22.38 & -22.48 & 2.216       & 1.923       \\     
J0216--0813 & 0.3317 & 333 $\pm$ 23 & 5.53  & 11.69       & 17.12   & -23.70 & -23.85 & 2.050       & 1.831       \\     
J0252+0039  & 0.2803 & 164 $\pm$ 12 & 4.40  & 11.25       & 18.15   & -22.24 & -22.35 & 0.950       & 0.878       \\     
J0330--0020 & 0.3507 & 212 $\pm$ 21 & 5.45  & 11.40       & 18.21   & -22.75 & -22.93 & 1.076       & 0.941       \\     
J0728+3835  & 0.2058 & 214 $\pm$ 11 & 4.21  & 11.30       & 16.83   & -22.73 & -22.84 & 1.567       & 1.348       \\     
J0737+3216  & 0.3223 & 338 $\pm$ 17 & 4.66  & 11.46       & 17.25   & -23.49 & -23.64 & 1.829       & 1.738       \\     
J0822+2652  & 0.2414 & 259 $\pm$ 15 & 4.45  & 11.38       & 17.10   & -22.89 & -23.00 & 1.644       & 1.458       \\     
J0841+3824  & 0.1159 & 225 $\pm$ 11 & 2.96  & 11.12       & 15.23   & -22.86 & -22.99 & 6.743       & 4.672       \\     
J0912+0029  & 0.1642 & 326 $\pm$ 16 & 4.58  & 11.60       & 15.77   & -23.22 & -23.34 & 3.034       & 2.452       \\     
J0935--0003 & 0.3475 & 396 $\pm$ 35 & 4.26  & 11.60       & 17.05   & -23.89 & -24.05 & 2.744       & 2.551       \\     
J0936+0913  & 0.1897 & 243 $\pm$ 12 & 3.45  & 11.17       & 16.62   & -22.74 & -22.86 & 1.876       & 1.691       \\     
J0946+1006  & 0.2219 & 263 $\pm$ 21 & 4.95  & 11.46       & 17.18   & -22.58 & -22.70 & 1.760       & 1.722       \\     
J0955+0101  & 0.1109 & 192 $\pm$ 13 & 1.83  & 10.83       & 17.04   & -20.89 & -21.02 & 1.549       & 0.840       \\     
J0956+5100  & 0.2405 & 334 $\pm$ 17 & 5.05  & 11.57       & 16.82   & -23.18 & -23.28 & 1.756       & 1.572       \\     
J0959+4416  & 0.2369 & 244 $\pm$ 19 & 3.61  & 11.23       & 17.12   & -22.82 & -22.93 & 1.500       & 1.381       \\     
J0959+0410  & 0.1260 & 197 $\pm$ 13 & 2.24  & 10.88       & 17.05   & -21.24 & -21.37 & 1.249       & 1.009       \\     
J1020+1122  & 0.2822 & 282 $\pm$ 18 & 5.12  & 11.54       & 17.47   & -22.94 & -23.05 & 1.156       & 1.037       \\     
J1023+4230  & 0.1912 & 242 $\pm$ 15 & 4.50  & 11.37       & 16.89   & -22.48 & -22.60 & 1.480       & 1.383       \\     
J1029+0420  & 0.1045 & 210 $\pm$ 11 & 1.92  & 10.78       & 16.24   & -21.66 & -21.77 & 1.771       & 1.193       \\     
J1032+5322  & 0.1334 & 296 $\pm$ 15 & 2.44  & 11.05       & 17.12   & -21.36 & -21.47 & 1.004       & 0.659       \\     
J1103+5322  & 0.1582 & 196 $\pm$ 12 & 2.78  & 10.98       & 16.63   & -22.27 & -22.39 & 1.927       & 1.217       \\     
J1106+5228  & 0.0955 & 262 $\pm$ 13 & 2.17  & 10.96       & 15.55   & -22.12 & -22.23 & 2.036       & 1.609       \\     
J1112+0826  & 0.2730 & 320 $\pm$ 20 & 6.19  & 11.65       & 17.41   & -22.90 & -23.02 & 1.160       & 1.010       \\     
J1134+6027  & 0.1528 & 239 $\pm$ 12 & 2.93  & 11.10       & 16.43   & -22.38 & -22.50 & 2.147       & 1.935       \\     
J1142+1001  & 0.2218 & 221 $\pm$ 22 & 3.52  & 11.22       & 17.13   & -22.65 & -22.76 & 1.779       & 1.640       \\     
J1143-0144  & 0.1060 & 269 $\pm$ 13 & 3.27  & 11.29       & 15.15   & -22.72 & -22.84 & 3.493       & 3.133       \\     
J1153+4612  & 0.1797 & 226 $\pm$ 15 & 3.18  & 11.05       & 17.25   & -21.97 & -22.08 & 1.037       & 1.029       \\     
J1204+0358  & 0.1644 & 267 $\pm$ 17 & 3.68  & 11.24       & 16.94   & -22.04 & -22.16 & 1.241       & 1.229       \\     
J1205+4910  & 0.2150 & 281 $\pm$ 14 & 4.27  & 11.40       & 16.81   & -22.88 & -23.00 & 1.977       & 1.706       \\     
J1213+6708  & 0.1229 & 292 $\pm$ 15 & 3.13  & 11.16       & 15.70   & -22.57 & -22.69 & 2.969       & 2.604       \\     
J1218+0830  & 0.1350 & 219 $\pm$ 11 & 3.47  & 11.21       & 15.89   & -22.61 & -22.72 & 2.739       & 2.414       \\     
J1250+0523  & 0.2318 & 252 $\pm$ 14 & 4.18  & 11.26       & 16.88   & -23.01 & -23.11 & 1.297       & 1.290       \\     
J1402+6321  & 0.2046 & 267 $\pm$ 17 & 4.53  & 11.46       & 16.52   & -23.02 & -23.14 & 2.251       & 1.997       \\     
J1403+0006  & 0.1888 & 213 $\pm$ 17 & 2.62  & 10.98       & 17.19   & -22.14 & -22.26 & 1.131       & 1.041       \\     
J1416+5136  & 0.2987 & 240 $\pm$ 25 & 6.08  & 11.56       & 17.71   & -22.85 & -22.97 & 1.227       & 1.082       \\     
J1420+6019  & 0.0629 & 205 $\pm$ 10 & 1.26  & 10.59       & 15.19   & -21.54 & -21.65 & 2.048       & 1.575       \\     
J1430+4105  & 0.2850 & 322 $\pm$ 32 & 6.53  & 11.73       & 16.96   & -23.49 & -23.58 & 1.728       & 1.668       \\     
J1432+6317  & 0.1230 & 199 $\pm$ 10 & 2.78  & 11.05       & 15.44   & -22.80 & -22.93 & 3.751       & 3.724       \\     
J1436--0000 & 0.2852 & 224 $\pm$ 17 & 4.80  & 11.36       & 17.41   & -23.03 & -23.13 & 1.776       & 1.587       \\     
J1443+0304  & 0.1338 & 209 $\pm$ 11 & 1.93  & 10.78       & 17.02   & -21.45 & -21.57 & 1.230       & 0.984       \\     
J1451--0239 & 0.1254 & 223 $\pm$ 14 & 2.33  & 10.92       & 16.08   & -22.21 & -22.33 & 2.167       & 2.010       \\     
J1525+3327  & 0.3583 & 264 $\pm$ 26 & 6.55  & 11.68       & 17.39   & -23.63 & -23.79 & 2.180       & 1.773       \\     
J1531--0105 & 0.1596 & 279 $\pm$ 14 & 4.71  & 11.43       & 15.95   & -22.92 & -23.04 & 2.573       & 2.201       \\     
J1538+5817  & 0.1428 & 189 $\pm$ 12 & 2.50  & 10.95       & 16.78   & -21.87 & -21.98 & 1.384       & 1.270       \\     
J1621+3931  & 0.2449 & 236 $\pm$ 20 & 4.97  & 11.47       & 16.95   & -23.09 & -23.22 & 1.908       & 1.698       \\     
J1627--0053 & 0.2076 & 290 $\pm$ 15 & 4.18  & 11.36       & 16.92   & -22.66 & -22.78 & 1.660       & 1.514       \\     
J1630+4520  & 0.2479 & 276 $\pm$ 16 & 6.91  & 11.69       & 17.00   & -23.06 & -23.18 & 1.537       & 1.394       \\     
J1636+4707  & 0.2282 & 231 $\pm$ 15 & 3.96  & 11.25       & 17.17   & -22.67 & -22.78 & 1.402       & 1.272       \\     
J2238--0754 & 0.1371 & 198 $\pm$ 11 & 3.08  & 11.11       & 16.33   & -22.18 & -22.30 & 1.963       & 1.748       \\     
J2300+0022  & 0.2285 & 279 $\pm$ 17 & 4.51  & 11.47       & 17.22   & -22.61 & -22.73 & 1.410       & 1.298       \\     
J2303+1422  & 0.1553 & 255 $\pm$ 16 & 4.35  & 11.42       & 16.07   & -22.69 & -22.83 & 2.591       & 2.108       \\     
J2321--0939 & 0.0819 & 249 $\pm$ 12 & 2.47  & 11.08       & 14.82   & -22.44 & -22.56 & 3.316       & 2.963       \\     
J2341+0000  & 0.1860 & 207 $\pm$ 13 & 4.50  & 11.35       & 16.48   & -22.81 & -22.94 & 2.661       & 2.078       \\     
\bottomrule
\end{tabular}
\flushleft
Notes: (1) Galaxy name. $(2)-(3)$ Galaxy redshift and SDSS-measured stellar velocity dispersion within the spectroscopic aperture of diameter 3 arcsec, both taken from T10, their Table~1. $(4)-(5)$ Einstein radius and total projected mass within a cylinder of radius equal to $\Rein$, taken from \citet{Auger.etal2009}, their Table~4. (6) \textit{I}-band apparent magnitude (\textit{F}814\textit{W}) derived from the MGE model (1$\sigma$ random error of 0.06 mag).
$(7)-(8)$ \textit{V} and \textit{r}-band absolute magnitudes (1$\sigma$ random error of 0.05 mag). (9) Major axis of the isophote containing half of the analytic total light of the MGE models (1$\sigma$ error of 10 per cent or 0.041 dex). (10) Circularized effective radius $\re =\sqrt{\Ae/\pi}$ where $\Ae$ is the area of the effective isophote containing half of the analytic total light of the MGE models (same error as $\remaj$).
\label{tab:mge}
\end{table*}

\begin{figure*}
\includegraphics[width=\linewidth]{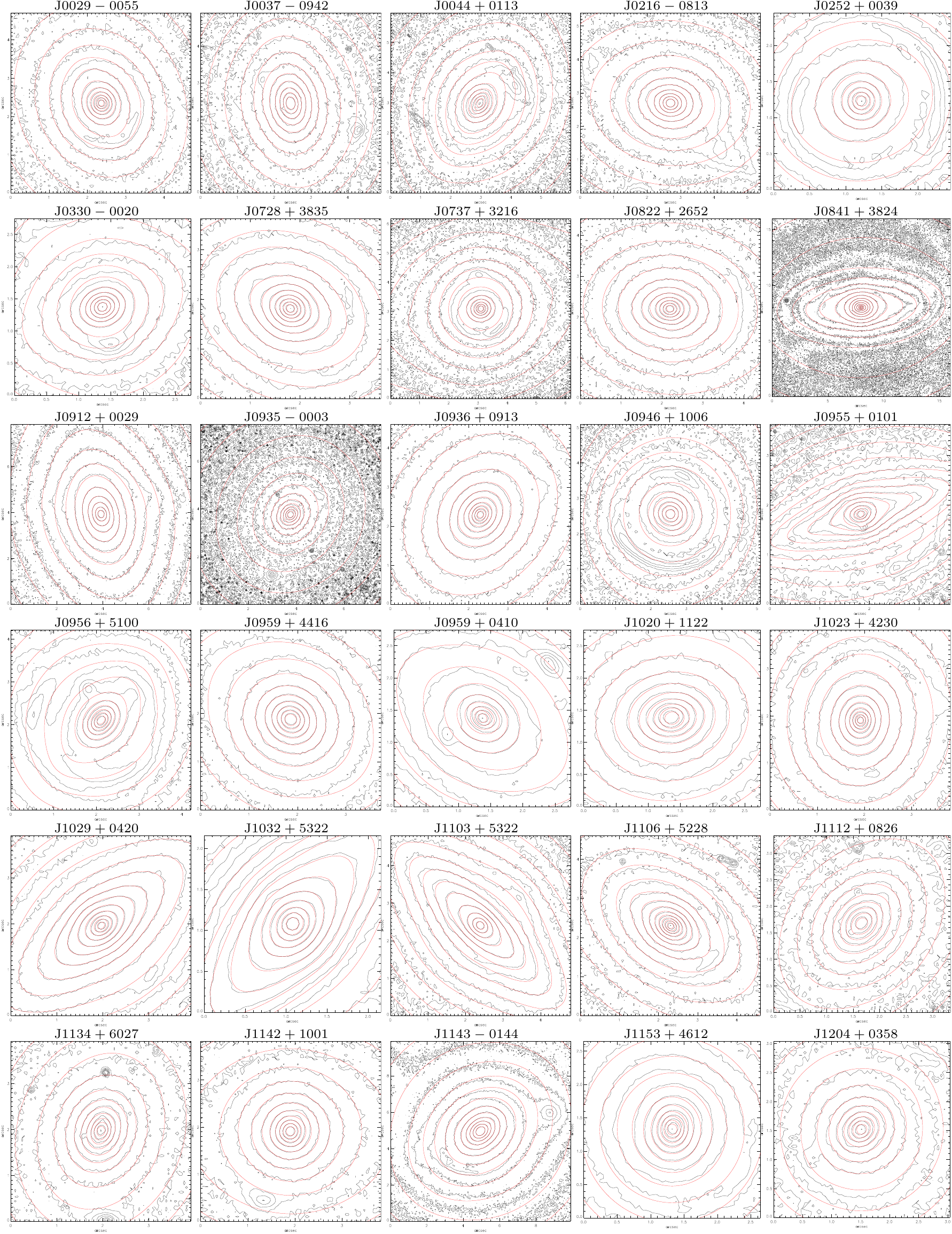}
\caption{Contour maps of the central regions ($\sim0.5\,\re$) of the WFC/\textit{F}814\textit{W} (\textit{I}-band) images of the 55 galaxies (black). The contours of the MGE surface
brightness, convolved with the proper PSF, are superimposed in red.}
\label{fig:MGE}
\end{figure*}
\begin{figure*}
\includegraphics[width=\linewidth]{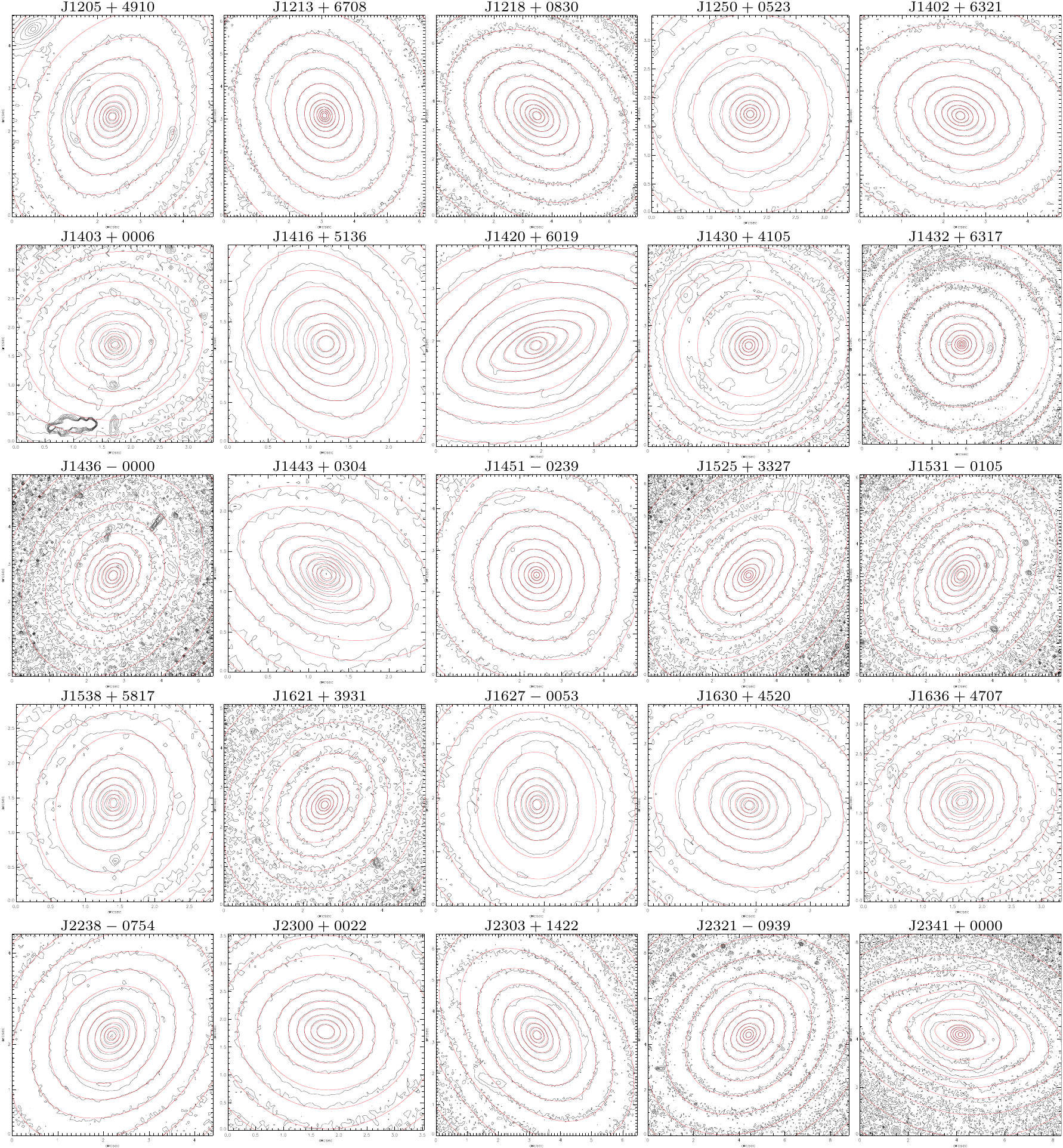}
\addtocounter{figure}{-1}
\caption{ -- \textit{continued}}
\end{figure*}

\begin{figure}
\includegraphics[width=\linewidth]{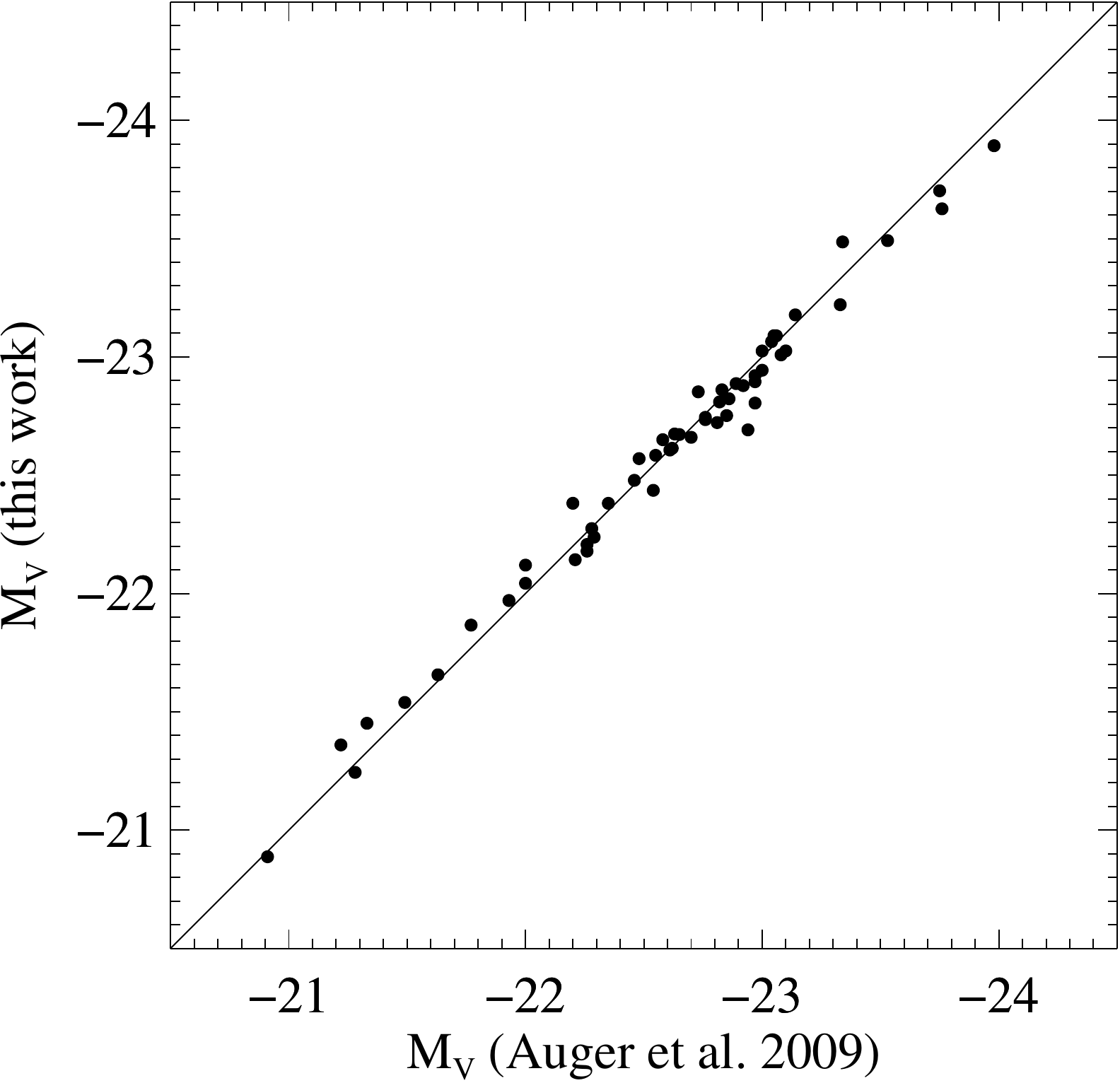}
\caption{\textit{V}-band absolute magnitudes for the SLACS sample, computed here and in \citet{Auger.etal2009}.}
\label{fig:MV_compare}
\end{figure}

For what concerns the DM halo density distribution, we adopt the NFW \citep{Navarro.etal1997} profile for two
main reasons.
The first is imposed by the few observational constraints at our disposal (see Section~\ref{sec:dyninfer}), which prevent us from exploring a more flexible DM halo profile, since the addition of further parameters to the models would make the problem completely undetermined. Thus, our results are valid under the assumption that the NFW profile is reliable in providing fair estimates for the DM fraction.
The second reason is that one of the motivations of this study is to investigate the possible bias introduced by T10
with the use of spherical isotropic Hernquist models to describe the stellar components of the SLACS sample, that is
apparently composed of non spherical galaxies (see Fig.\ref{fig:MGE}). Thus, in order to disentangle the effects
produced by this approximation, we make use of the same DM density profile adopted by T10, that is the untruncated
NFW profile
\begin{equation}
\rho_{\rm h}(r)=\dfrac{\rho_{\rm crit}~\delta_{\rm c}r_{\rm h}}{r\left(1+r/r_{\rm h}\right)^2},
\label{eq:NFW}
\end{equation}
with fixed $r_{\rm h}=30$ kpc.
We then perform a one-dimensional MGE fit to Eq.~(\ref{eq:NFW}) in order to recover the DM surface density in units of $\Msun$ pc$^{-2}$, and add the DM halo to the dynamical modelling (Sect.\ref{sec:jam}).

Finally, we apply a similar procedure for the BH, parametrizing it with a single gaussian with a dispersion of 0.01 arcsec.
The BH mass is chosen adopting the $M_{\rm BH}-\se$ relation of \citet{Gultekin.2009} for elliptical galaxies, where, for each galaxy, $\se$ (i.e., the luminosity averaged stellar velocity dispersion within $\re$) is computed starting from $\sigma_*$ (the SDSS-measured velocity dispersion, luminosity-averaged within a circular aperture of radius 1.5 arcsec) and using the conversion formula in eq.~(1) of \citet{Cappellari.etal2006}, thus accounting for aperture correction.

\subsubsection{The stellar kinematics}\label{sec:jam}
The model velocity fields are computed using the Jeans anisotropic MGE (JAM) modelling method of \citet{Cappellari.2008}, which can be applied to an axisymmetric stellar distribution, described by a three-integral distribution function. 
This method assumes a velocity ellipsoid aligned with the cylindrical coordinates $(R,z,\varphi)$, and a constant
vertical anisotropy parametrized by $\beta_{z}=1-\sigma^2_z/\sigma^2_R$.
For our models we fix $\beta_{z}=0.2$, which has been found to be representative of local ETGs \citep{Cappellari.etal2007}. However, relaxing this assumption, and considering isotropic models ($\beta_{z}=0$) as done in T10, negligibly affects our results.
Moreover, for simplicity we assume a spatially constant $M_{*}/L$, even if recent studies found evidences for a IMF
dependence on galactocentric distance \citep[e.g.,][]{Martin-Navarro.etal2014, Pastorello.etal2014}. 
These evidences do not make our results invalid, since this assumption simply implies that our measured $M_{*}/L$
represents a mean value in the observed region (which typically has size $r\lesssim \re$), as already done by
\citet{Cappellari.etal2013XV}. This does not exclude, for example, that the IMF might be universal in the outer disc
components and vary only within bulges or spheroids (see e.g., \citealt{Dutton.etal2013a}).

The main ingredients of the dynamical modelling are the galaxy surface brightness in units of $L_{\odot}\,$pc$^{-2}$, and the galaxy surface density of the total mass distribution in units of $\Msun$ pc$^{-2}$. This last is the sum of the three components (stars, DM and BH) obtained as described in Sect.~\ref{sec:mass_struc}, where the stellar one is multiplied by a stellar mass-to-light ratio $\MLdyn$ that is the quantity we want to retrieve (as will be explained in Sect.~\ref{sec:dyninfer}).
Then the only free parameters left are $\beta_{z}$ and the inclination angle $i$, whose values have to be provided or assumed.
Indeed, once the MGE parametrization of the surface brightness profile is obtained, the MGE parametrization of the intrinsic light profile can be easily and analytically recovered for a choice of the inclination angle $i$. Here we adopt $i=60$\textdegree, i.e., the average inclination for random orientations, and, whenever the axial ratio of the gaussians does not allow deprojection for this inclination, we adopt the minimum inclination permitted. 
Note that a significant error in the adopted value of $i$ would produce errors smaller that 10 per cent on the retrieved mass-to-light ratio, if the observed axial ratio is $q<0.7$ (see \citealt{Cappellari.etal2006} Fig.~4 for a detailed discussion).
Given these inputs, with the JAM$^1$ method we are able to directly compute the projected second velocity moment along
the line-of-sight (LOS) $V_{\rm rms}$, with a single numerical quadrature.
Finally, in order to compare $V_{\rm rms}$ with $\sigma_*$, we convolve it with a gaussian PSF with a FWHM of 1.5 arcsec, as typical for SDSS observations, and then we compute a luminosity-weighted average inside the 3 arcsec diameter SDSS fiber.

\begin{table*}
\caption{Mass-to-light ratios and $\fDM$ of the models for the 55 galaxy SLACS subsample}
\begin{tabular}{ccccc}
\toprule   
Name        & $\log\MLself$             & $\log\MLdyn$              & $\log\MLpop$                 & $\fDM$                \\
            & [$\Msun/\Lsunr$]          & [$\Msun/\Lsunr$]          & [$\Msun/\Lsunr$]             &                       \\
(1)         & (2)                       & (3)                       & (4)                          & (5)                  \\
 \midrule                                                                                                                                                                       
J0029--0055  & 0.693 & 0.609 & 0.610     & $<0.207$                    \\    
J0037--0942  & 0.665 & 0.648 & 0.615     & 0.036   \\     
J0044+0113   & 0.771 & 0.708 & 0.642     & $<0.174$                    \\     
J0216--0813  & 0.747 & 0.737 & 0.635     & $<0.401$                    \\     
J0252+0039   & 0.439 &-0.178 & 0.646     & 0.804   \\     
J0330--0020  & 0.487 & 0.279 & 0.557     & 0.350   \\     
J0728+3835   & 0.491 & 0.331 & 0.598     & 0.416   \\     
J0737+3216   & 0.819 & 0.662 & 0.572     & $<0.061$                    \\     
J0822+2652   & 0.681 & 0.665 & 0.531     & 0.041   \\     
J0841+3824   & 0.688 & 0.672 & 0.724     & 0.079   \\     
J0912+0029   & 0.872 & 0.848 & 0.728     & $<0.174$                    \\     
J0935--0003  & 0.923 & 0.784 & 0.545     & $<0.511$                    \\     
J0936+0913   & 0.640 & 0.611 & 0.600     & $<0.123$                    \\     
J0946+1006   & 0.871 & 0.824 & 0.632     & 0.137   \\     
J0955+0101   & 0.858 & 0.846 & 0.640     & $<0.246$                    \\     
J0956+5100   & 0.826 & 0.765 & 0.646     & $<0.070$                    \\     
J0959+4416   & 0.631 & 0.609 & 0.524     & 0.003   \\     
J0959+0410   & 0.777 & 0.763 & 0.661     & 0.024   \\     
J1020+1122   & 0.662 & 0.595 & 0.634     & 0.194   \\     
J1023+4230   & 0.674 & 0.578 & 0.662     & 0.256   \\     
J1029+0420   & 0.672 & 0.581 & 0.637     & $<0.044$                    \\     
J1032+5322   & 1.004 & 0.804 & 0.670     & $<0.009$                    \\     
J1103+5322   & 0.599 & 0.539 & 0.690     & $<0.246$                    \\     
J1106+5228   & 0.668 & 0.593 & 0.680     & $<0.027$                    \\     
J1112+0826   & 0.793 & 0.783 & 0.625     & 0.001   \\     
J1134+6027   & 0.714 & 0.698 & 0.677     & 0.001   \\     
J1142+1001   & 0.624 & 0.437 & 0.665     & 0.338   \\     
J1143-0144   & 0.754 & 0.734 & 0.713     & 0.060   \\     
J1153+4612   & 0.678 & 0.656 & 0.658     & 0.001   \\     
J1204+0358   & 0.835 & 0.805 & 0.677     & $<0.065$                    \\     
J1205+4910   & 0.785 & 0.751 & 0.667     & $<0.086$                    \\     
J1213+6708   & 0.817 & 0.717 & 0.684     & $<0.022$                    \\     
J1218+0830   & 0.640 & 0.555 & 0.667     & 0.294   \\     
J1250+0523   & 0.548 & 0.509 & 0.522     & $<0.078$                    \\     
J1402+6321   & 0.706 & 0.620 & 0.661     & 0.256   \\     
J1403+0006   & 0.656 & 0.607 & 0.468     & $<0.180$                    \\     
J1416+5136   & 0.612 & 0.458 & 0.575     & 0.265   \\     
J1420+6019   & 0.596 & 0.513 & 0.649     & $<0.055$                    \\     
J1430+4105   & 0.774 & 0.755 & 0.640     & 0.044   \\     
J1432+6317   & 0.569 &-0.112 & 0.658     & 0.862   \\     
J1436--0000  & 0.588 & 0.440 & 0.652     & 0.338   \\     
J1443+0304   & 0.686 & 0.647 & 0.594     & $<0.040$                    \\     
J1451--0239  & 0.679 & 0.635 & 0.579     & $<0.146$                    \\     
J1525+3327   & 0.622 & 0.433 & 0.612     & 0.478   \\     
J1531--0105  & 0.728 & 0.720 & 0.706     & $<0.103$                    \\     
J1538+5817   & 0.577 & 0.429 & 0.671     & 0.256   \\     
J1621+3931   & 0.571 & 0.365 & 0.629     & 0.431   \\     
J1627--0053  & 0.830 & 0.766 & 0.653     & $<0.065$                    \\     
J1630+4520   & 0.671 & 0.555 & 0.685     & 0.275   \\     
J1636+4707   & 0.620 & 0.608 & 0.639     & 0.030   \\     
J2238--0754  & 0.621 & 0.461 & 0.631     & 0.375   \\     
J2300+0022   & 0.804 & 0.754 & 0.661     & 0.115   \\     
J2303+1422   & 0.779 & 0.744 & 0.691     & 0.147   \\     
J2321--0939  & 0.709 & 0.700 & 0.657     & 0.001   \\     
J2341+0000   & 0.616 & 0.219 & 0.470     & 0.676   \\     
\bottomrule
\end{tabular}
\flushleft
Notes: (1) Galaxy name. 
(2) Total mass-to-light ratio of the mass-follows-light dynamical models (Sect.~\ref{sec:dynmodel}) in the \textit{r}-band (1$\sigma$ error of 14 per cent or 0.056 dex). 
$(3)-(4)$ \textit{r}-band stellar mass-to light ratios derived from the dynamical and the stellar population synthesis models, respectively (Sects.~\ref{sec:dynmodel} and ~\ref{sec:spsmodel}). The 1$\sigma$ error in $\MLdyn$ is 28 per cent (0.106 dex), and 7 per cent (0.03 dex) for $\MLpop$.
(5) DM fraction enclosed within a sphere of radius $\re$, derived from the dynamical models (1$\sigma$ error of 0.16).
\label{tab:mge2}
\end{table*}

\subsubsection{Inferring the parameters of the dynamical models}\label{sec:dyninfer}
For each galaxy in the sample we built a set of galaxy models, whose mass structure and kinematical configuration have been already described in Sec.~\ref{sec:mass_struc} and \ref{sec:jam}, respectively. We then use two observationally derived quantities to constrain the best model: the SDSS-measured aperture stellar velocity dispersion $\sigma_{*}$, provided by the SDSS database, and the total projected mass $\Mein$ enclosed within the Einstein ring of radius $\Rein$, calculated by \citet{Auger.etal2009}. These quantities are reproduced in Table~\ref{tab:mge} with their errors; for $\Mein$ we adopt an error of 5 per cent.

Within a set, the models have the same values for $(M_{\rm BH},i,\beta_{\rm z})$, and they differ only in the mass normalization of the two main components: the stellar population and the DM halo.
In practice, we choose a sufficiently wide range within which the $r$-band stellar mass-to-light ratio $\MLdyn$ is allowed to vary, and we multiply the MGE model surface density by $\MLdyn$; in this way we convert the MGE model into a mass density. Analogously, we choose a range for the DM mass normalization by using the parameter $\fDM$, i.e., the DM fraction within a sphere of radius equal to one effective radius $\re$; obviously $0\leq\fDM\leq1$. We then build a model for each couple of values ($\MLdyn$, $\fDM$), and we choose the best-fitting model by means of chi-squared minimization on the two observables ($\sigma_{*}$, $\Mein$). 
The chi-square maps for the whole sample (Fig.~\ref{fig:chi2}) show some degeneracy between $\fDM$ and $\MLdyn$.
In general, the DM fraction is very low ($\fDM\lesssim 0.4$) and for nearly half of the sample it tends to zero, probably due to systematics. Few galaxies are indeed scarcely reproduced by the NFW profile here adopted, probably due to systematic errors associated to $\sigma_*$, or to difficulties in the retrieval of the MGE parametrization because of strong lens disturbances, or to the lensing analysis.
Table~\ref{tab:mge2} shows the best-fitting $\MLdyn$ and $\fDM$, and reports the associated typical errors. These are the median values of the 1$\sigma$ errors, computed projecting the white areas in Fig.~\ref{fig:chi2} in the allowed region of the parameters.

Isotropic models (not shown here) result overall in lower $\fDM$ and higher $\MLdyn$, which do not affect any of our results.
The dynamical contribution of the BH is irrelevant, since removing it from the modelling results in a negligible increase of $\MLdyn$: the percentage variation has a median value of 1 per cent for the whole sample, and is always smaller than 7 per cent. 

Finally, we also built another set of dynamical models where the total mass distribution follows that of the light (mass-follows-light models).
These are less sophisticated dynamical models that are constrained only using $\sigma_*$, and whose best-fitting mass-to-light ratios are reported in Table~\ref{tab:mge2} as $\MLself$.
The associated typical error is reported in Table~\ref{tab:mge2}. It is computed propagating the errors in $\sigma_*$ and in the JAM modelling (6 per cent, as evaluated in \citealt{Cappellari.etal2006} from a wider exploration of dynamical modelling approaches), and adopting the median value for all the galaxies.

\begin{figure*}
\includegraphics[width=\linewidth]{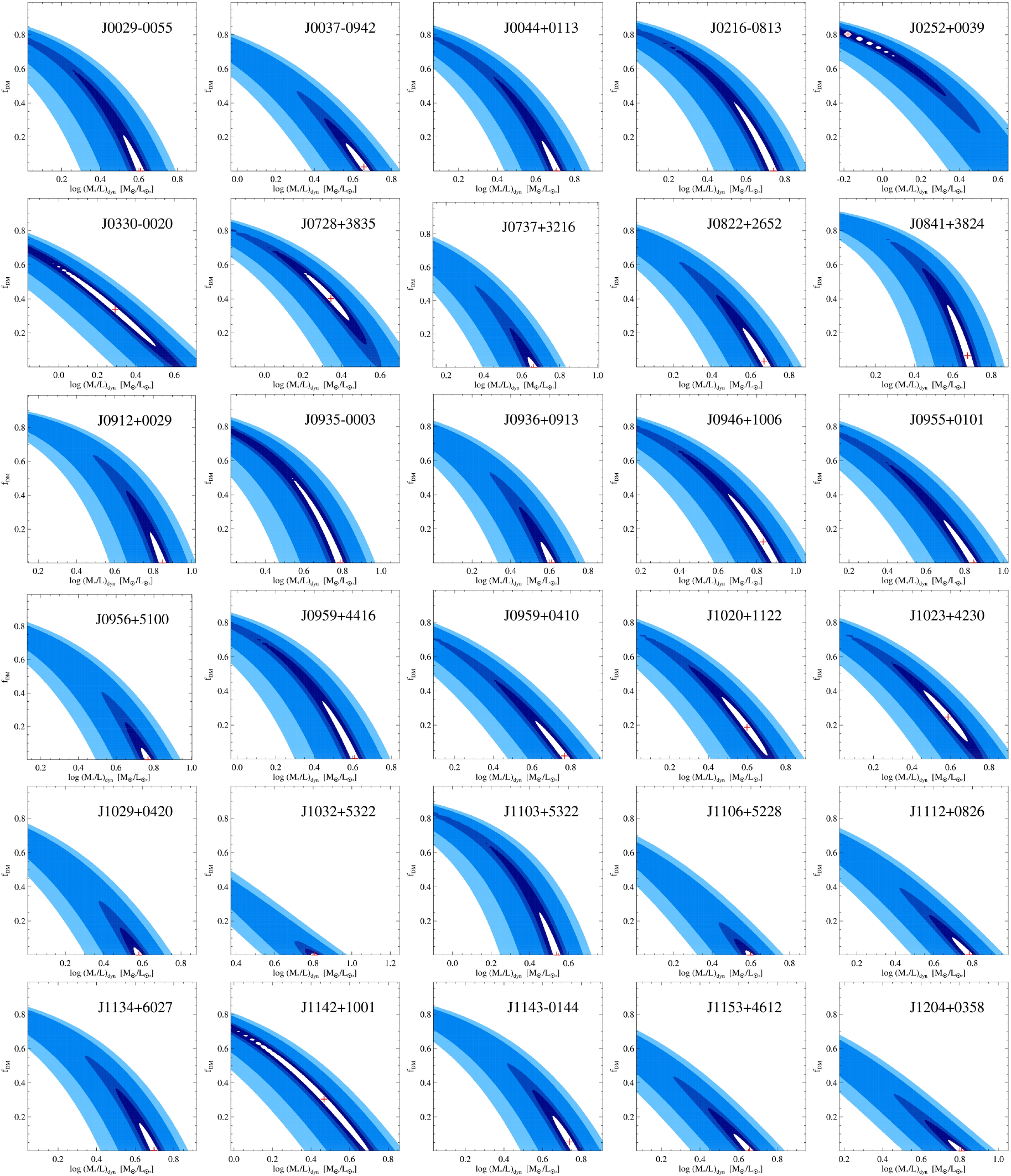}
\caption{$\Delta\chi^2$ contour maps obtained from the dynamical models, as a function of the DM fraction $\fDM$ (vertical axis) and $r$-band stellar mass-to light ratio $\MLdyn$ (horizontal axis). The red cross locates the minimum chi-square value. The $1,2,3\,\sigma$ confidence levels for 1 degree of freedom ($\Delta\chi^2=1,4,9)$ are shown in white, dark blue and light blue, respectively.}
\label{fig:chi2}
\end{figure*}

\begin{figure*}
\includegraphics[width=\linewidth]{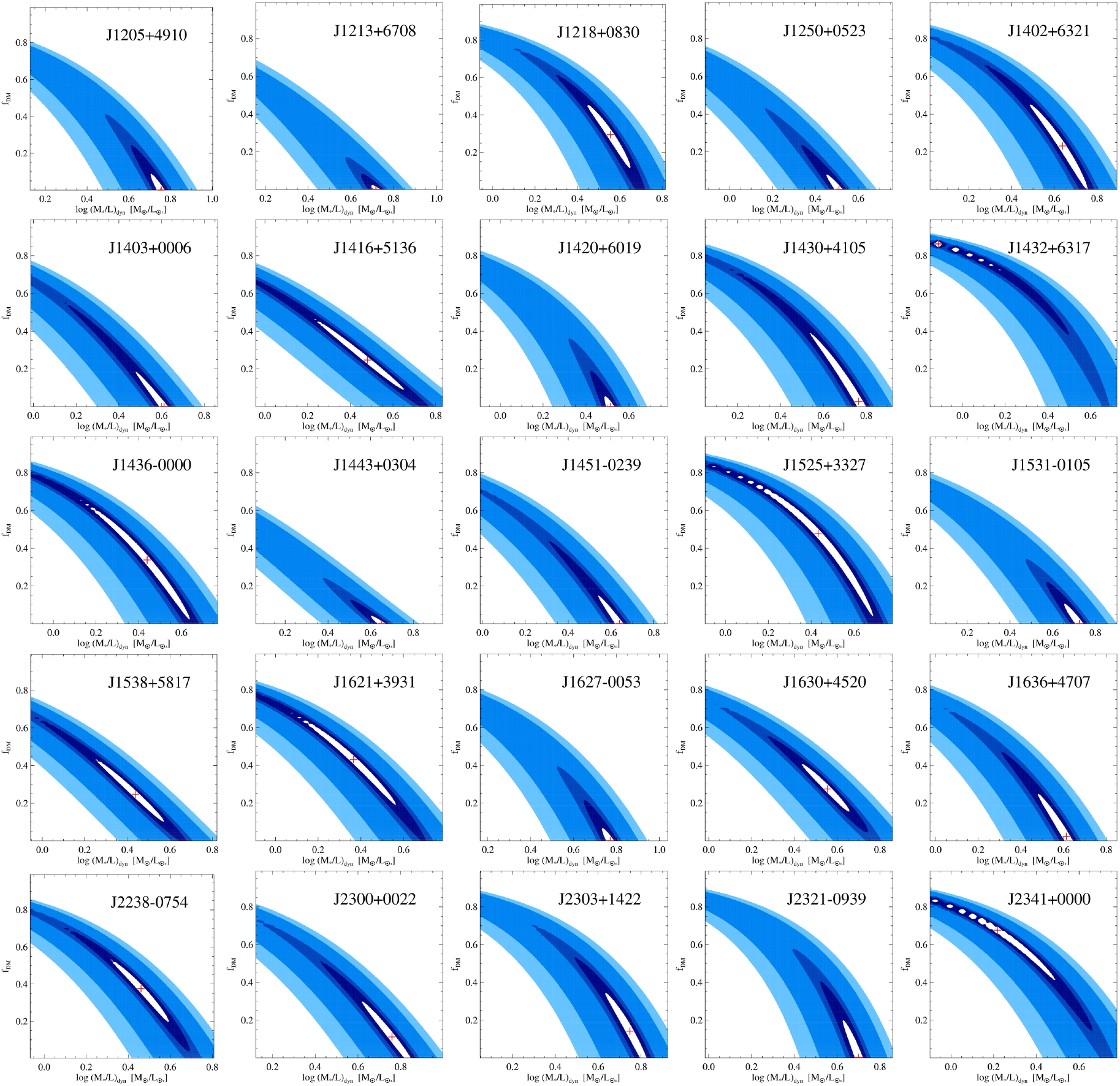}
\addtocounter{figure}{-1}
\caption{ -- \textit{continued}}
\end{figure*}

\subsection{The stellar population synthesis modelling} \label{sec:spsmodel}
Our stellar population synthesis models are performed applying a full-spectrum fitting approach to SDSS spectra, and using a selection of the simple stellar population (SSP) models of \citet{Vazdekis.etal2010}\footnote{Available at http://miles.iac.es/}, which are based on the MILES stellar spectral library \citep{SanchezBlazquez.etal2006}, and cover the wavelength range $3540-7410$ \AA~at 2.50 \AA~(FWHM) spectral resolution. 
In particular, we adopt the \citet{Salpeter.1955} IMF as reference, and we select the MILES SSP models with age $t\geqslant 1$ Gyr and metallicity $-1.71\leqslant[M/H]\leqslant0.22$: this leads to a total of 156 SSPs with 26 logarithmically-spaced ages, and metallicity values $[M/H] = [-1.71, -1.31, -0.71, -0.40, 0.00, 0.22]$.
For each galaxy then, the spectral fitting is allowed to use only SSPs with age not greater than the age of the Universe at the galaxy redshift, reducing the number of SSP templates to $N<156$.
The full-spectrum fitting is performed with the \textsc{ppxf} software$^1$, which implements the Penalized Pixel-Fitting
method of \citet{Cappellari.etal2004}, and, for each galaxy, returns the best fitting matrix of weights $w$ (to be
multiplied by the SSP templates). Then, the stellar mass-to-light ratio in the \textit{r}-band associated to the
population model is
\begin{equation}
\MLpop=\dfrac{\sum^N_{j=1} w_j\, M_j^{\rm nogas}}{\sum^N_{j=1} w_j\, L_{j,\,r}},
\end{equation}
where $M_j^{\rm nogas}$ and $L_{j,\,r}$ are the stellar mass (including neutron stars and black holes, but excluding the gas lost by the stars during stellar evolution) and the $r$-band luminosity of the $j$-th SSP, respectively.
In general, for these un-regularized fits, we find that $N\leqslant5$, and in most of the cases $N=2$
with the older and more metal rich SSP having $w\simeq1$.

The spectral fitting has been performed also using the ppxf keyword REGUL: in this way the fitting procedure is forced to apply a linear regularization to the weights (see equation 18.5.10 of \citealt{Press.etal1992}), obtaining a smoother solution than the unregularized fit. The regularized solution is as statistically good as the unregularized one, being still consistent with the observations, but it is more physically plausible and representative of the galaxy population since it reduces the scatter in the retrieved population parameters (i.e., age and metallicity) of the solution.
The regularized $\MLpop$ slightly underestimate the unregularized ones by 0.02 dex, with an rms scatter of 0.014 dex; this would imply errors of 7 per cent in the individual $\MLpop$.
Finally we find that our results are robust against plausible variations of the REGUL parameter, so that here we present the results obtained with the regularized solutions.
In both fits we make use of a 10-th degree multiplicative Legendre polynomial to correct the continuum shape for calibration effects and account for possible intrinsic dust absorption.
The best-fitting $\MLpop$ are reported in Table~\ref{tab:mge2} for each galaxy.

\section{Results}\label{sec:results}
Here we present our main results regarding the mass-to-light ratios we derived and their correlation with the stellar velocity dispersion ($M/L-\sigma$ relation), and we focus mostly on the implications concerning the IMF normalization.

\subsection{Mass-follows-light models}
\begin{figure}
\includegraphics[width=\linewidth]{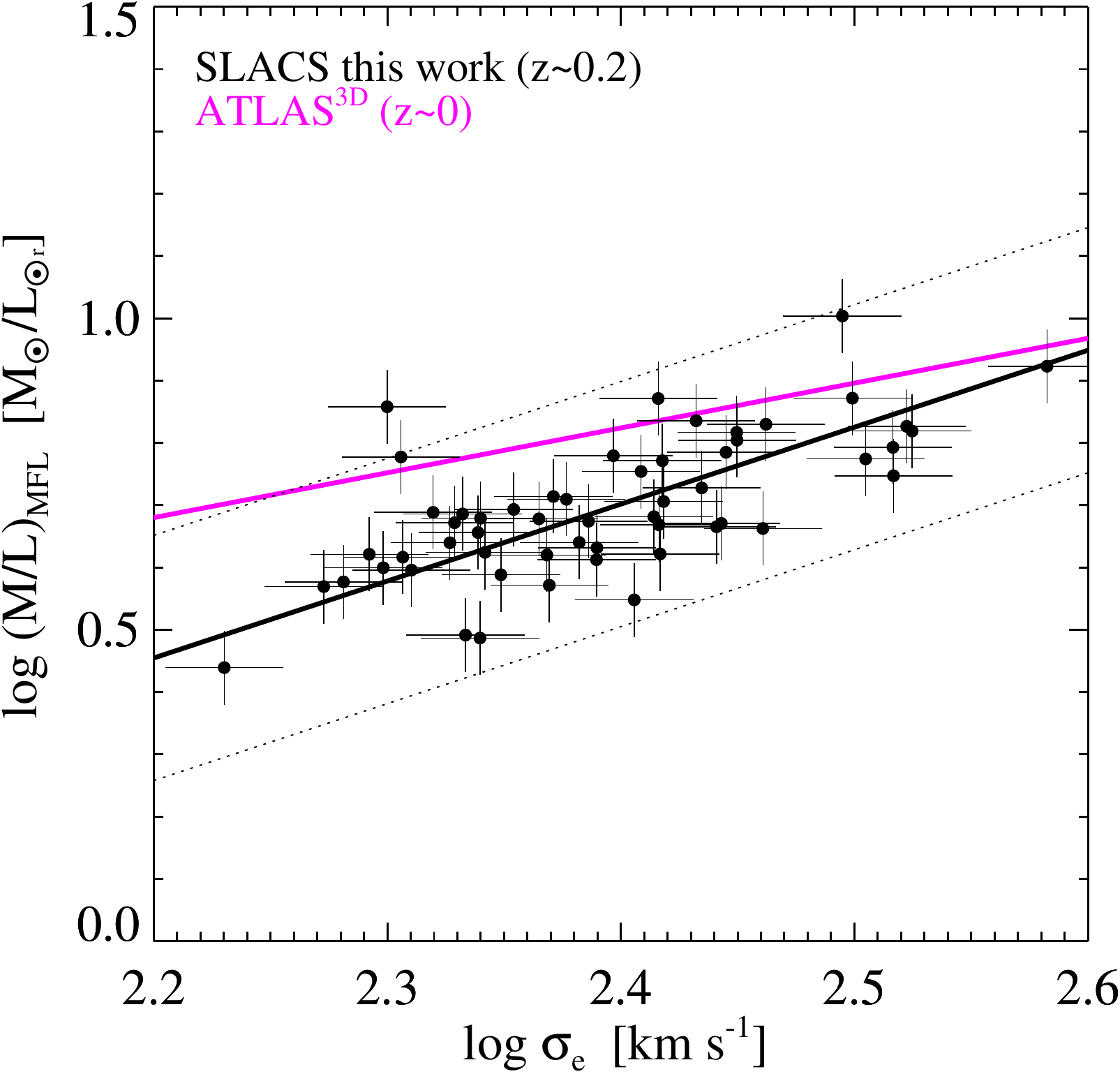}
\caption{\textit{r}-band mass-to-light ratios of the mass-follows-light dynamical models for the SLACS sample, as a function of $\se$, and their best-fitting relation (black line). The magenta line is the best-fitting relation for the \ATLAS sample \citep{Cappellari.etal2013XV}.
The values of $\se$ are computed as described at the end of Sect.~\ref{sec:mass_struc}.
Our best-fitting relation is obtained with \textsc{lts$\_$linefit}, and the dotted lines mark the 3$\sigma$ bands (enclosing 99.7\% of the values for a Gaussian distribution). Outliers deviating more than 3$\sigma$ from the best-fitting relation were automatically excluded from the fit (i.e., points beyond the dotted lines).}
\label{fig:ML_sigma}
\end{figure}
We recall that these models have a total mass profile that follows that of the light, and are tuned to reproduce only the galaxy surface brightness and the SDSS-measured aperture velocity dispersion.
Figure~\ref{fig:ML_sigma} shows the mass-to-light ratios $\MLself$ we derived for these dynamical models as a function of $\se$, and the black line is the $M/L-\se$ relation we obtained for the SLACS sample. The relation is of the form 
\begin{equation}
\log(M/L)_{r} = (1.24\pm0.14)\times\log\left(\frac{\se}{200\,\kms}\right)+ (0.58 \pm0.02), 
\label{eq:MLse}
\end{equation}
and has an rms scatter of 0.08 dex. The best-fitting relation has been obtained using the \textsc{lts$\_$linefit}
routine$^1$ of \citet{Cappellari.etal2013XV}, which allows and fits for intrinsic scatter, and robustly manages the
presence of outliers. 
In the fit we consider a typical error of 6 per cent for $\se$, and we quadratically co-added JAM modelling errors of 14 per cent, plus distance errors, plus 5 per cent errors for our photometry.
When compared with previous similar estimates for different samples of ETGs, local and not (e.g., \citealt{Cappellari.etal2006,vanderMarel.etal2007,Cappellari.etal2013XV}), our relation is slightly steeper. 
For example, analogous mass-follows-light models have been built also for the \ATLAS \citep{Cappellari.etal2013XV} and SAURON samples \citep{Cappellari.etal2006}, leading to $M/L-\se$ relations shallower than our, and with higher zero-points (e.g., see the magenta line in Fig.~\ref{fig:ML_sigma}). The \ATLAS sample consists of local galaxies, while the SLACS galaxies reside at higher redshifts (the median redshift for the SLACS sample is $z\simeq0.2$), so that their stellar populations are younger on average, resulting in lower stellar mass-to-light ratios. Indeed the offset between the two samples can be accounted just by considering passive evolution. For reference, a solar metallicity ($[Z/H]=0$) passively evolving stellar population varies its ${M/L}_r$ by $\sim0.10$ dex from an age of 11 Gyr to 14 Gyr ($z\sim0.2$ to $z=0$, assuming it formed at $z_{\rm form}=\infty$), according to the models of \citet{Maraston.2005}. This value provides a lower limit to the expected passive ${M/L}_r$ 
variation we should observe.

A possible explanation for the steeper slope, instead, could be provided by indications that the $M/L-\se$ relation
might steepen at the high $\se$ end \citep{Zaritsky.etal2006}. In fact, the SLACS sample consists mostly of high
velocity dispersion galaxies ($200\,\kms\lesssim\se\lesssim400\,\kms$) due to its selection criteria, while for
example the volume limited \ATLAS sample extends from high-intermediate $\se$ galaxies to very low $\se$ systems
($50\,\kms\lesssim\se\lesssim250\,\kms$). Thus the two relations shown in Fig.~\ref{fig:ML_sigma} have been
obtained sampling different ranges in velocity dispersion, that barely intersect each other.

\subsection{Dependency of the IMF normalization on velocity dispersion}
\begin{figure}
\includegraphics[width=\linewidth]{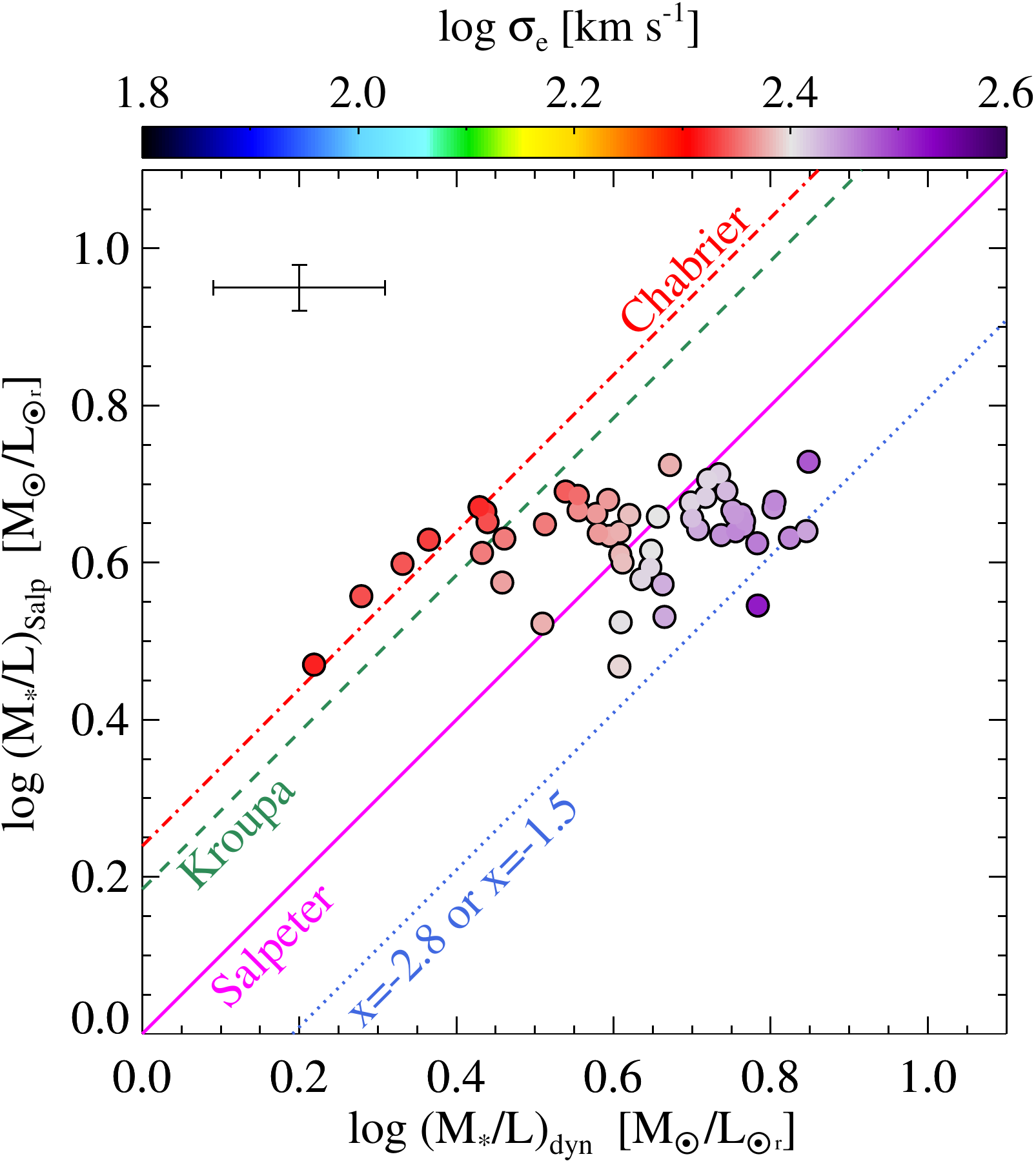}
\caption{The stellar mass-to-light ratios $\MLpop$ for a Salpeter IMF (Sect.~\ref{sec:spsmodel}) are shown as a function of the dynamical stellar mass-to-light ratios $\MLdyn$ (Sect.~\ref{sec:dynmodel}), both derived in the \textit{r}-band. The colours of the symbols code the galaxy velocity dispersion: in place of the individual $\se$ values, here we show the two-dimensional LOESS smoothed $\se$ values (see the top colour bar). A representative error bar is shown at the top-left.
Two galaxies resulting in too high and unrealistic DM fractions (J0252+0039 and J1432+6317) have been excluded from the plot.
The diagonal lines are computed from the \citet{Vazdekis.etal2010} models for a population with solar metallicity.}
\label{fig:MLdyn_vs_pop}
\end{figure}

Figure~\ref{fig:MLdyn_vs_pop} shows the two sets of stellar mass-to-light ratios obtained from our dynamical and stellar population synthesis models, one against the other. Note that, at variance with $\MLself$, the dynamical mass-to-light ratios $\MLdyn$  here shown are purely stellar, since a NFW DM halo has been included in the modelling, so that they can be directly compared with $\MLpop$. Thus, if for example the IMF of ETGs is universal and Salpeter-like, $\MLdyn$ should be very similar to $\MLpop$, which has been calculated under this assumption (i.e., all galaxies should lie close to the magenta line, with some scatter). If otherwise ETGs have a lighter IMF, like Chabrier or Kroupa, one would expect to find that $\MLpop$ systematically overestimates $\MLdyn$ by the same percentage, for the whole sample. 
The situation apparent in Fig.~\ref{fig:MLdyn_vs_pop} is somewhat different: galaxies do not lie near one of the lines representing different IMFs, but are distributed across all of them. The scatter is significant compared to the typical error, and reveals that some galaxies are actually more properly represented by a lighter or a heavier IMF normalization. 
This suggests a variation of the IMF for ETGs, that seems also to correlate with the galaxy velocity dispersion, with low-$\se$ galaxies being consistent with a Chabrier or Kroupa-like IMF, while medium and high-$\se$ galaxies agree with a Salpeter or heavier IMF.
Note that our results are equally consistent with both a bottom heavy and top heavy IMF trend \citep[as considered
by][]{Weidner.etal2013}, since the approach we use does not constrain the shape of the IMF directly, but only the
overall mass normalization.
In Fig.~\ref{fig:MLdyn_vs_pop} each galaxy is coloured according to its LOESS-smoothed value of $\se$, as done in
\citet{Cappellari.etal2013XX} (their fig. 11). Applying the LOESS$^1$ method \citep{Cleveland.1979}, we
evaluated mean values of $\se$ that are the result of an average over the neighbouring galaxies, weighted with the
relative distances. In this way, one aims to reconstruct the average values characterizing the underlying galaxy
population, i.e., the values one should expect to obtain when disposing of much larger galaxy samples.

\begin{figure}
\includegraphics[width=\linewidth]{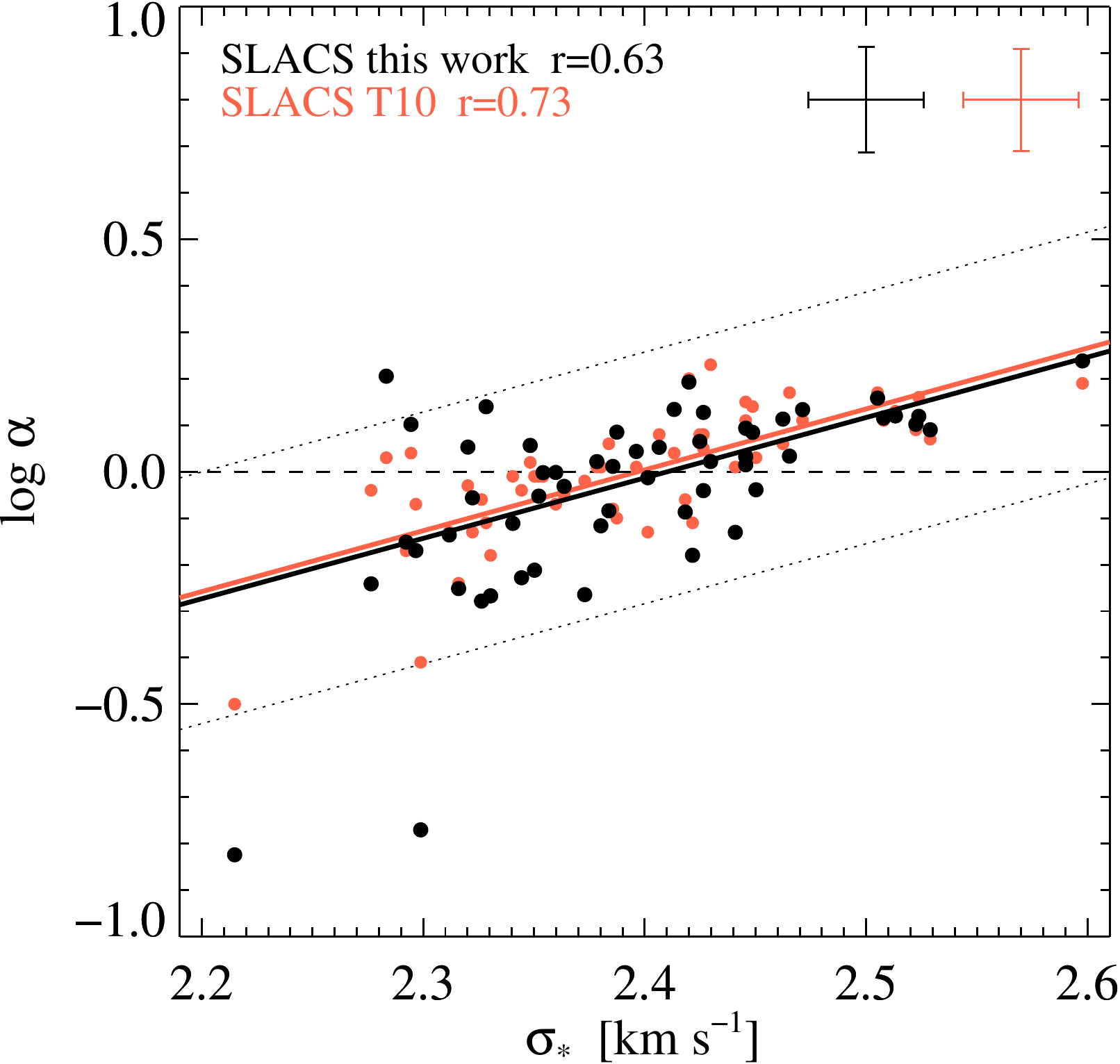}
\caption{IMF mismatch parameter $\alpha$ as a function of $\sigma_*$ for the SLACS sample.
Red points are taken from T10, as well as their best-fitting relation represented by the red line.
Black points refer to the values computed in this work. The black line is our best-fitting relation obtained with \textsc{lts$\_$linefit}, and the dotted lines mark the 3$\sigma$ bands (enclosing 99.7\% of the values for a Gaussian distribution). Outliers deviating more than 3$\sigma$ from the best-fitting relation were automatically excluded from the fit (i.e., points beyond the dotted lines). The value of the linear correlation coefficient $r$ is also reported. Representative error bars are shown at the top-right: for the data of T10 we compute the median error.}
\label{fig:IMF_mismatch}
\end{figure}
\begin{figure}
\includegraphics[width=\linewidth]{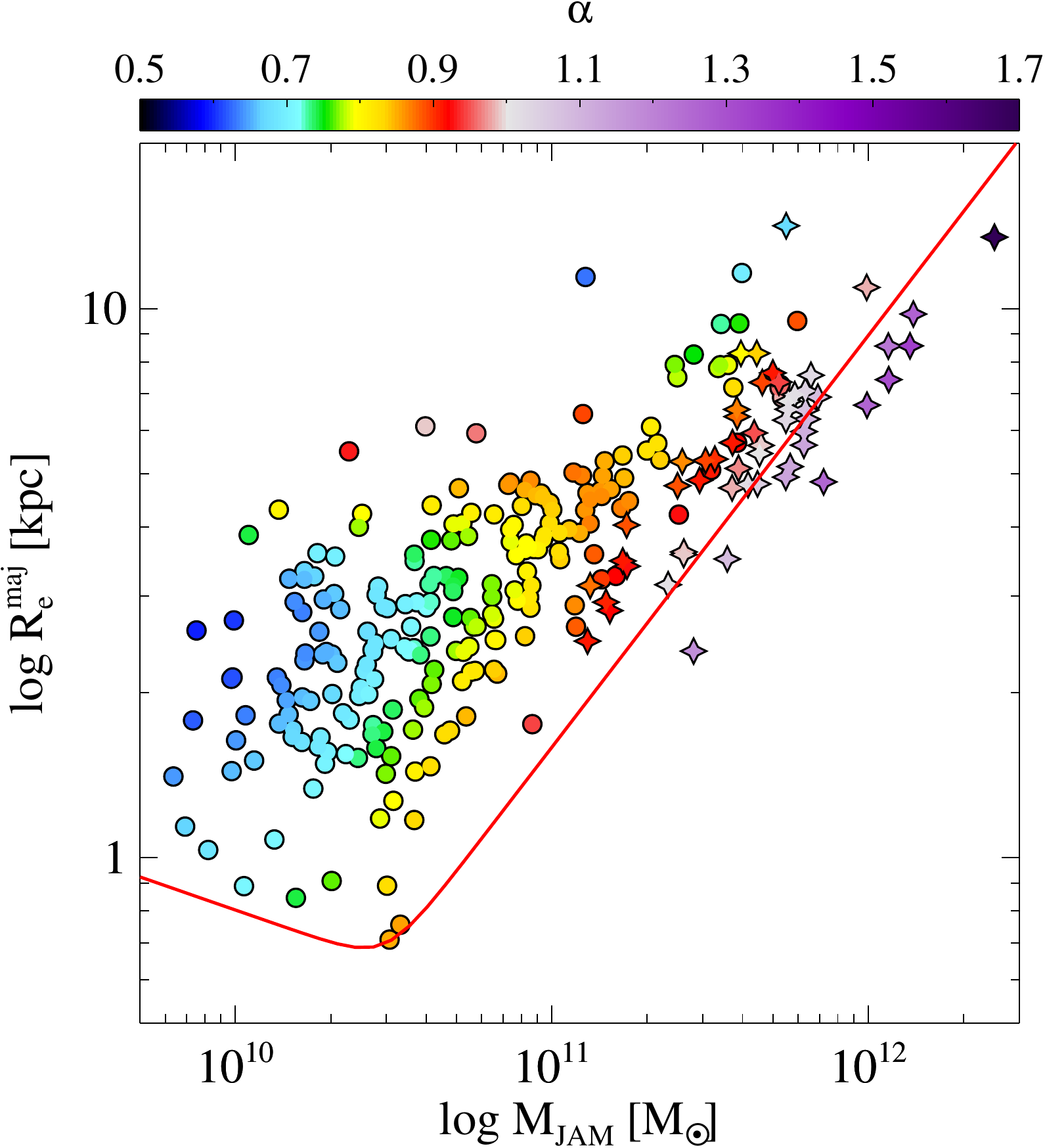}
\caption{$\remaj$, the major axis of the isophotes containing half of the analytic total light of the MGE models, is shown as function of $\Mjam$, the total mass of the mass-follows-light models (i.e. $\Mjam=\MLself\times L_r$), for the SLACS (stars) and \ATLAS samples (circles). The colours of the symbols code the ratio $\alpha=\MLdyn/\MLpop$: in place of the individual $\alpha$ values, here we show the two-dimensional LOESS smoothed $\alpha$ values (see the top colour bar).
The red line shows the zone of exclusion relation given by equation (4) of \citet{Cappellari.etal2013XX}, for the \ATLAS sample.}
\label{fig:mass_size}
\end{figure}

Another way of seeing this variation is by looking at the IMF mismatch parameter $\alpha\equiv\MLdyn/\MLpop$. Figure~\ref{fig:IMF_mismatch} shows the logarithm of $\alpha$ as a function of $\sigma_*$, as already done in T10 (see their fig.~4, central panel). Here, the red points refer to the values obtained by T10, while our results are shown in black, and the solid lines are the respective best-fitting relations. Note that the dynamical models of T10 consist of spherical isotropic models, with a stellar component following a \citet{Hernquist.1990} profile. Moreover their stellar population synthesis models were built using multicolour \textit{HST} photometry, while ours are based on full-spectrum fitting. Regardless of the very different approaches adopted, we find that the two works produce essentially the same result pointing toward an IMF variation, with high-$\sigma_*$ galaxies being consistent on average with a Salpeter normalization. Our relation is
\begin{equation}
\log\alpha =(1.3\pm0.23)\times\log\left(\frac{\sigma_*}{200\,\kms}\right)-(0.14 \pm0.03),
\label{eq:alphas*}
\end{equation}
with an rms scatter of 0.1 dex; in the fit we consider a median error of 6 per cent for $\sigma_*$, and we quadratically co-added the dynamical modelling errors of 28 per cent, plus distance errors, plus population models errors of 7 per cent, plus 5 per cent errors for our photometry.
Our relation is very similar to that reported in T10. However, inspecting Fig.~\ref{fig:IMF_mismatch} a difference must be noted: our dynamical modelling produces a weaker correlation, in the sense that our points are more scattered in the $(\log\alpha,\sigma_*)$ plane with respect to T10 points. This is reasonably due to the use of a more flexible parametrization of the light profiles. 
Indeed, given its nature, the SLACS sample is likely to include also compact galaxies, and using a density profile with a fixed internal slope (like the Hernquist profile) to fit all the galaxies might artificially produce some correlation, by overestimating the stellar mass in the high-$\se$ compact galaxies.
Figure~\ref{fig:mass_size} illustrates the type of galaxies that are in the $\sigma$-selected SLACS sample (stars), compared to the volume-selected \ATLAS sample (circles): it can be noticed that they are quite massive and dense, since they fill the lower envelope of the galaxy distribution in the $(\remaj,\Mjam)$ at the high mass end (i.e., with the smaller $\remaj$ for $\Mjam>10^{11}\Msun $).

\begin{figure}
\includegraphics[width=\linewidth]{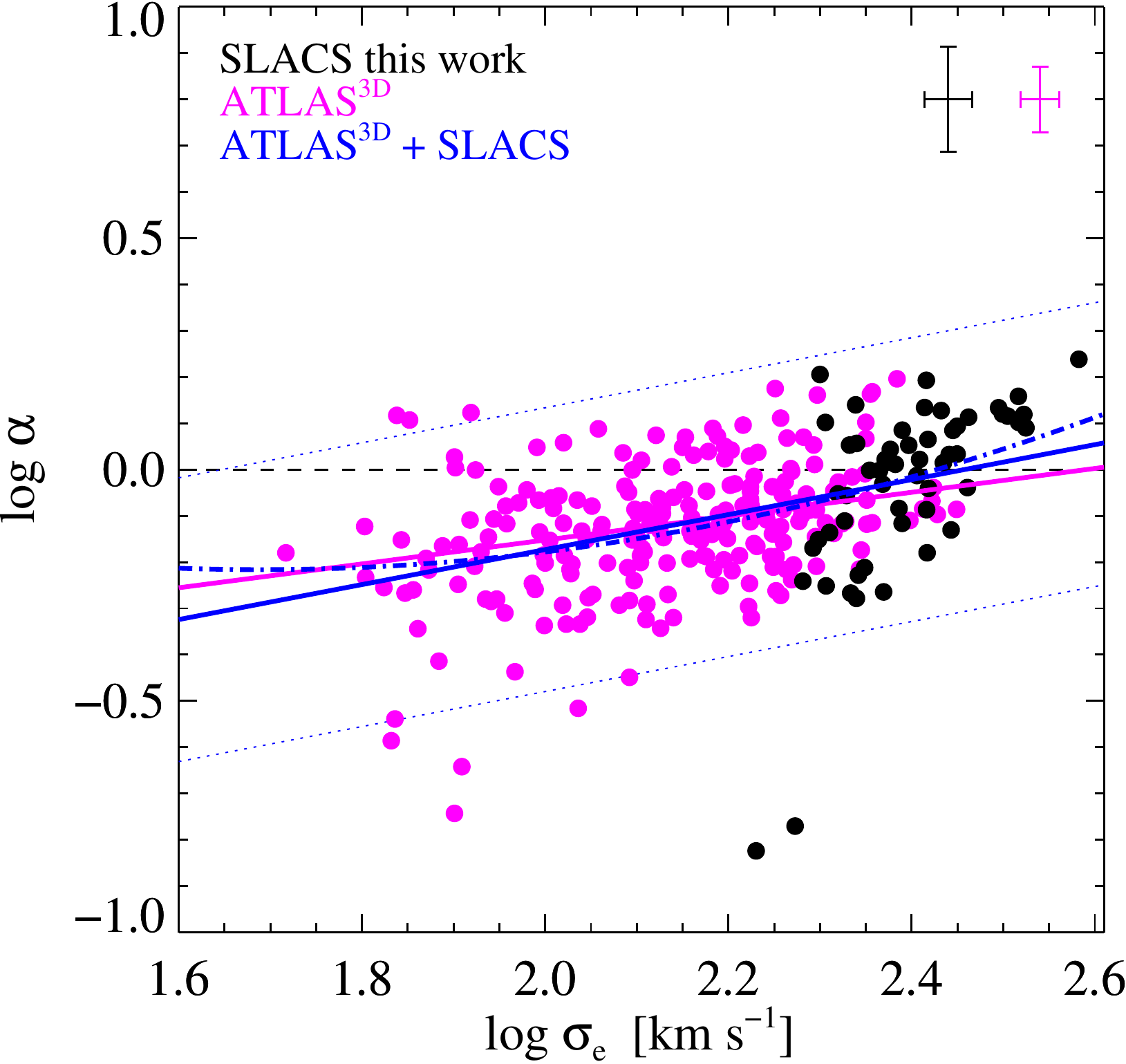}
\caption{The IMF mismatch parameter $\alpha$ is shown as a function of $\se$ for the SLACS (black) and the \ATLAS sample (magenta). The magenta solid line is the best-fitting relation for the subset of the whole \ATLAS sample made of 223 galaxies with the stellar absorption line-strength index H$\beta<2.3$~\AA, taken from \citet{Cappellari.etal2013XX}. The blue solid line is the best-fitting relation for the two samples put together, obtained with \textsc{lts$\_$linefit}, and the blue dotted lines mark the 3$\sigma$ bands (enclosing 99.7\% of the values for a Gaussian distribution). Outliers deviating more than 3$\sigma$ from the best-fitting relation were automatically excluded from the fit (i.e., points beyond the dotted lines). The blue dot-dashed line is a parabolic fit to both samples together, performed with the \textsc{mpfitfun} routine. Representative error bars are shown at the top-right: for the data of \citet{Cappellari.etal2013XX} we compute the median error.}
\label{fig:IMF_mismatch_together}
\end{figure}
Finally, the analysis we conducted on the SLACS sample is analogous to the one performed on the \ATLAS sample, both in terms of the dynamical and stellar population approach. This allows us to directly compare the respective results, and merge the two samples homogeneously analysed to infer some global insights on the IMF of ETGs.
Fig.~\ref{fig:IMF_mismatch_together} shows the IMF mismatch parameter as a function of $\se$ for the two samples (i.e., SLACS in black and \ATLAS in magenta). Here, notwithstanding the heterogeneity of the samples in terms of selection criteria, galaxy redshift and mass range, one can immediately appreciate how the black points seem to follow the same relation of the magenta points, but extending to higher $\se$ values. Indeed, the magenta solid line, representing the best-fitting relation for the \ATLAS sample, is only slightly shallower that the blue solid line, obtained by fitting both samples together; in particular, we find for the whole sample SLACS$\,+\,$\ATLAS
\begin{equation}
\log\alpha =(0.38\pm0.04)\times\log\left(\frac{\se}{200\,\kms}\right)+(-0.06\pm0.01), 
\label{eq:alphase_all}
\end{equation}
with an rms scatter of 0.12 dex. The similarity of the two best-fitting relations is even more remarkable when comparing them with the steeper relation we find for the SLACS sample alone (Eq.~\ref{eq:alphas*}). Note that the steepness of the slope in Eq.~\ref{eq:alphas*} is not due to the fact that $\alpha$ is fitted as a function of $\sigma_*$ instead of $\se$, since we find a very similar result also for $\se$ (slightly steeper).
This shows that the slope of the $\alpha-\se$ relation is very sensitive to the $\se$ range, with a considerable increase for $\se\gtrsim250\,\kms$, and suggests that the relation is not a simple single power law.
In this scenario, the steepness of the $\alpha-\se$ correlation, found by T10 and confirmed here, for the ETGs of the SLACS sample is a natural consequence of the velocity dispersion selection nature of the SLACS sample.
We then try to fit a parabola to the whole sample SLACS$\,+\,$\ATLAS, obtaining
\begin{equation}
\begin{split}      
\log\alpha =&(0.40\pm0.15)\times\log\left(\frac{\se}{200\,\kms}\right)^2+\\
&(0.49\pm0.05)\times\log\left(\frac{\se}{200\,\kms}\right)+(-0.07\pm0.01),
\end{split}
\label{eq:parabola}
\end{equation}
with an rms scatter of 0.12 dex.

Thus, by homogeneously studying ETGs collected over a very wide and unprecedented range of $\se$ and $M_*$, we have provided a comprehensive insight about the IMF normalizaton for this morphological type of galaxies, showing that the IMF gets heavier for increasing $\se$, and becomes Salpeter-like at $\se\simeq250\,\kms$.
The issue of the IMF variability for the \ATLAS sample has also been studied by \citet{Tortora.etal2014} within
the MOND framework, obtaining results consistent with the ones from Newtonian dynamics plus DM.

\section{Discussion and Conclusions} \label{sec:conclusions}
In this work we studied the mass normalization of the IMF of ETGs, exploiting information derived from gravitational lensing, stellar dynamics and stellar population synthesis models, and making use of high-quality photometric and spectroscopic data.
We selected 55 ETGs belonging to the SLACS sample and constructed dynamical and stellar population synthesis models for each galaxy.
Our dynamical models are built solving the Jeans axisymmetric anisotropic equations with the JAM method of \citet{Cappellari.2008}; they reproduce in detail the \textit{HST} galaxy images and are constrained using the SDSS-measured velocity dispersion and the mass within the Einstein radius. Our stellar population synthesis models are computed with the full-spectrum fitting technique and are based on the SSP models of \citet{Vazdekis.etal2010}. 
We derived accurate estimates of stellar mass-to-light ratios from the two sets of models, $\MLdyn$ and $\MLpop$ respectively.

From the comparison of the two estimates of stellar mass-to-light ratios, we find a trend of IMF with velocity
dispersion, where, on average, the IMF normalization smoothly varies from Kroupa/Chabrier for galaxies with
$\se\sim 90\,\kms$, up to a bottom-heavy Salpeter-like IMF for galaxies with $\se\sim270\,\kms$
(Fig.~\ref{fig:MLdyn_vs_pop}). 
This change of IMF normalization as a function of $\se$ is significant beyond the extent of the error estimates in the stellar the mass-to-light ratios, and thus suggests an intrinsic systematic variation of the stellar IMF for ETGs.

With our accurate and realistic modelling of the stellar profiles, our analysis provides an improvement over the study of T10, conducted on the same ETG sample. Notwithstanding the different and independent approaches adopted, we confirm their finding of a steep correlation between the IMF mismatch parameter $\alpha=\MLdyn/\MLpop$ and the galaxy velocity dispersion (Fig.~\ref{fig:IMF_mismatch}); however our relation has a slightly lower correlation coefficient, presumably due to relaxing the restrictive assumption of a fixed stellar density profile to fit the whole galaxy sample.

We also built mass-follows-light dynamical models and computed total mass-to-light ratios $\MLself$ for them. We find a $\MLself-\se$ correlation steeper than previous analogous estimates for different local ETG samples (e.g., the $\MLself-\se$ relation for the 260 ETGs \ATLAS sample), and with a lower zero-point (Fig.~\ref{fig:ML_sigma}). 
The SLACS sample resides at higher redshift and is likely to include galaxies with younger stellar populations; indeed the offset in the zero-points can be accounted for by passive evolution between $z\sim0.2$ and $z=0$.
The different slope instead could be an effect of the different $\se$ range spanned by the samples, in accordance with \citet{Zaritsky.etal2006} that suggests a steepening this relation as a function of $\se$.
Note that the slope of the $\MLself-\se$ relation gives an upper limit to any systematic increase of the IMF mass normalization with $\se$.

Finally, as an important outcome of analysing the SLACS galaxies with a procedure that is homogeneous with that adopted for the \ATLAS galaxies \citep{Cappellari.etal2013XX},
we could merge the two samples. In this way, we explored the behaviour of ETGs in the $\alpha-\se$ plane with the largest sample ever, where ETGs of all $\se$ values from $50\,\kms$ to $\sim350\,\kms$ are well represented.
We found that the volume-limited \ATLAS sample and the velocity dispersion selected SLACS galaxies smoothly merge in a unique shallower relation in the $(\alpha,\se)$ plane (Fig.~\ref{fig:IMF_mismatch_together}).
From this comprehensive analysis, we find that the $\alpha-\se$ relation might not be linear, and that the
slope inferred  depends on the range of $\se$ covered by the galaxies. This is significantly different for the \ATLAS
(volume seected) and SLACS sample (velocity dispersion selected).

\section*{Acknowledgements}
We thank the anonymous referee for the useful comments and suggestions which helped to improve this paper.
S.P. acknowledges Léon V. E. Koopmans and Iary Davidzon for useful discussions.
M.C. acknowledges support from a Royal Society University Research Fellowship.
T.T. acknowledges support from the Packard Foundation in the form of a Packard Research Fellowship.
L.C., S.P. and S.P. are supported by the PRIN MIUR 2010-2011, project "The Chemical and Dynamical Evolution of the Milky Way and Local Group Galaxies", prot. 2010LY5N2T.

\bibliographystyle{mn2e}
\bibliography{references.bib}

\appendix

\section{MGE model parameters}
\label{app:mge}

For each galaxy, the parameters of the best-fitting MGE parametrizations
of the projected light are presented in Table~\ref{tab:mgeee}.

\begin{table*}
\caption{MGE parameters for the deconvoled $r$-band surface brightness.}
\label{tab:mgeee}
\begin{tabular*}{\linewidth}{c@{\hspace{1.em}}c@{\hspace{1.4em}}c@{\hspace{1.6em}}@{\hspace{3em}}c@{\hspace{1.em}}c@{\hspace{1.4em}}c@{\hspace{1.6em}}@{\hspace{3em}}c@{\hspace{1.em}}c@{\hspace{1.4em}}c@{\hspace{1.6em}}@{\hspace{3em}}c@{\hspace{1.em}}c@{\hspace{1.4em}}c@{\hspace{1.6em}}}
\toprule
$\log I_i$ & $\log\sigma_i$ & $q_i$ & $\log I_i$ & $\log\sigma_i$ & $q_i$ & $\log I_i$ & $\log\sigma_i$ & $q_i$ & $\log I_i$ & $\log\sigma_i$ & $q_i$\\
$[L_{\odot\,r}\,\mathrm{pc}^{-2}]$ & [arcsec] & & $[L_{\odot\,r}\,\mathrm{pc}^{-2}]$ & [arcsec] & & $[L_{\odot\,r}\,\mathrm{pc}^{-2}]$ & [arcsec] & & $[L_{\odot\,r}\,\mathrm{pc}^{-2}]$ & [arcsec] &\\[1pt]
\midrule
\multicolumn{12}{c}{} \\
\multicolumn{3}{c@{\hspace{3em}}}{J0029--0055}& \multicolumn{3}{c@{\hspace{3em}}}{J0037--0942}& \multicolumn{3}{c@{\hspace{3em}}}{J0044+0113} & \multicolumn{3}{c@{\hspace{3em}}}{J0216--0813} \\
\\
  3.895 &  --1.532 &    0.922 &   3.124 &  --1.323 &    0.891  &    3.552 &  --1.264 &    0.693 &    3.392 &  --1.532 &    0.842 \\   
  3.700 &  --1.050 &    0.941 &   3.442 &  --0.924 &    0.693  &    3.735 &  --0.786 &    0.567 &    3.536 &  --1.006 &    0.842 \\   
  3.372 &  --0.684 &    0.903 &   3.494 &  --0.773 &    0.693  &    3.507 &  --0.565 &    0.820 &    3.454 &  --0.762 &    0.842 \\   
  2.714 &  --0.183 &    0.792 &   3.355 &  --0.496 &    0.693  &    2.549 &  --0.209 &    0.554 &    3.183 &  --0.450 &    0.794 \\   
  1.902 &  --0.076 &    0.941 &   3.068 &  --0.212 &    0.693  &    2.871 &  --0.113 &    0.842 &    2.718 &  --0.126 &    0.792 \\   
  2.081 &    0.174 &    0.792 &   2.225 &    0.127 &    0.891  &    2.422 &    0.193 &    0.842 &    2.377 &    0.171 &    0.792 \\   
  1.677 &    0.570 &    0.838 &   2.356 &    0.152 &    0.693  &    1.521 &    0.304 &    0.297 &    1.935 &    0.531 &    0.842 \\   
        &          &          &   1.833 &    0.593 &    0.891  &    1.018 &    0.658 &    0.297 &          &          &          \\   
        &          &          &         &          &           &    1.820 &    0.658 &    0.758 &          &          &          \\  
\\                                                                                                                                                    
\multicolumn{3}{c@{\hspace{3em}}}{J0252+0039}&\multicolumn{3}{c@{\hspace{3em}}}{J0330--0020}& \multicolumn{3}{c@{\hspace{3em}}}{J0728+3835}& \multicolumn{3}{c@{\hspace{3em}}}{J0737+3216} \\
\\
  4.005 &  --1.532 &    0.941 &   4.247 &  --1.532 &    0.787  &   4.156  &  --1.532 &    0.845 &    3.631 &  --1.532 &    0.941 \\
  3.462 &  --1.078 &    0.941 &   3.215 &  --1.094 &    0.745  &   3.811  &  --1.069 &    0.852 &    3.778 &  --1.119 &    0.982 \\
  3.240 &  --0.724 &    0.941 &   3.670 &  --0.925 &    0.829  &   3.583  &  --0.790 &    0.838 &    3.474 &  --0.700 &    0.900 \\
  2.199 &  --0.143 &    0.652 &   3.028 &  --0.601 &    0.842  &   3.125  &  --0.521 &    0.941 &    2.880 &  --0.188 &    0.990 \\
  2.684 &  --0.136 &    0.941 &   2.679 &  --0.406 &    0.743  &   2.670  &  --0.303 &    0.512 &    2.317 &    0.142 &    0.842 \\
  1.834 &    0.212 &    0.941 &   2.529 &  --0.111 &    0.743  &   2.698  &  --0.044 &    0.746 &    1.815 &    0.552 &    0.842 \\
        &          &          &   1.976 &  --0.055 &    0.842  &   2.089  &    0.183 &    0.792 &          &          &          \\
        &          &          &   1.949 &    0.300 &    0.835  &   1.484  &    0.541 &    0.495 &          &          &          \\
        &          &          &         &          &           &   1.568  &    0.541 &    0.804 &          &          &          \\                                                                                                                                      
\\
 \multicolumn{3}{c@{\hspace{3em}}}{J0822+2652}& \multicolumn{3}{c@{\hspace{3em}}}{J0841+3824}&\multicolumn{3}{c@{\hspace{3em}}}{J0912+0029}& \multicolumn{3}{c@{\hspace{3em}}}{J0935--0003} \\
\\
  3.935 &  --1.532 &    0.792 &    4.318 &  --1.532 &    0.760 &    3.349 &  --1.433 &    0.801 &   3.459 &  --1.162 &    0.862  \\    
  3.676 &  --1.074 &    0.792 &    3.810 &  --0.996 &    0.990 &    3.464 &  --0.917 &    0.870 &   3.499 &  --0.848 &    0.822  \\    
  3.536 &  --0.796 &    0.792 &    3.056 &  --0.551 &    0.531 &    3.097 &  --0.629 &    0.732 &   3.276 &  --0.570 &    0.823  \\    
  3.173 &  --0.470 &    0.743 &    3.296 &  --0.423 &    0.792 &    3.172 &  --0.410 &    0.727 &   2.819 &  --0.247 &    0.862  \\    
  2.713 &  --0.118 &    0.792 &    2.878 &  --0.142 &    0.848 &    2.520 &  --0.092 &    0.565 &   2.371 &    0.058 &    0.862  \\    
  2.186 &    0.177 &    0.759 &    2.202 &    0.278 &    0.446 &    2.680 &    0.029 &    0.597 &   2.129 &    0.532 &    0.862  \\    
  1.791 &    0.528 &    0.743 &    2.249 &    0.493 &    0.446 &    2.329 &    0.047 &    0.941 &         &          &           \\    
        &          &          &    1.598 &    0.937 &    0.798 &    2.120 &    0.392 &    0.841 &         &          &           \\    
        &          &          &          &          &          &    2.053 &    0.427 &    0.443 &         &          &           \\    
        &          &          &          &          &          &    1.584 &    0.826 &    0.657 &         &          &           \\    
\\
\multicolumn{3}{c@{\hspace{3em}}}{J0936+0913} & \multicolumn{3}{c@{\hspace{3em}}}{J0946+1006}& \multicolumn{3}{c@{\hspace{3em}}}{J0955+0101}&\multicolumn{3}{c@{\hspace{3em}}}{J0956+5100} \\
\\
  4.075 &  --1.532 &    0.822 &    3.338 &  --1.532 &    0.990 &    4.135 &  --1.532 &    0.720 &   3.818 &  --1.532 &    0.743  \\  
  3.780 &  --1.019 &    0.818 &    3.374 &  --0.940 &    0.990 &    3.634 &  --1.064 &    0.599 &   3.827 &  --1.025 &    0.743  \\  
  3.437 &  --0.800 &    0.826 &    3.250 &  --0.568 &    0.990 &    3.286 &  --0.772 &    0.842 &   3.496 &  --0.751 &    0.743  \\  
  3.233 &  --0.506 &    0.828 &    2.587 &  --0.023 &    0.990 &    2.841 &  --0.520 &    0.842 &   3.236 &  --0.504 &    0.743  \\  
  2.574 &  --0.270 &    0.842 &    1.475 &    0.287 &    0.743 &    2.878 &  --0.036 &    0.248 &   2.679 &  --0.024 &    0.743  \\  
  2.467 &  --0.060 &    0.792 &    1.642 &    0.585 &    0.743 &    2.459 &    0.116 &    0.411 &   2.353 &    0.040 &    0.941  \\  
  2.342 &    0.198 &    0.817 &          &          &          &    1.324 &    0.396 &    0.842 &   1.760 &    0.268 &    0.743  \\  
  1.706 &    0.584 &    0.833 &          &          &          &    1.630 &    0.396 &    0.442 &   1.791 &    0.579 &    0.743  \\  
        &          &          &          &          &          &          &          &          &         &          &           \\
        &          &          &          &          &          &          &          &          &         &          &           \\
\\
\multicolumn{3}{c@{\hspace{3em}}}{J0959+4416}& \multicolumn{3}{c@{\hspace{3em}}}{J0959+0410} & \multicolumn{3}{c@{\hspace{3em}}}{J1020+1122}& \multicolumn{3}{c@{\hspace{3em}}}{J1023+4230} \\
\\
3.272 &  --1.253 &    0.941  &    4.098 &  --1.532 &    0.801 &    3.731 &  --1.413 &    0.792 &    4.322 &  --1.532 &    0.866 \\
3.593 &  --1.018 &    0.941  &    3.523 &  --1.123 &    0.857 &    3.811 &  --0.967 &    0.792 &    3.781 &  --1.026 &    0.891 \\
3.429 &  --0.748 &    0.877  &    3.466 &  --0.866 &    0.746 &    3.444 &  --0.656 &    0.792 &    3.230 &  --0.754 &    0.842 \\
3.079 &  --0.492 &    0.892  &    3.062 &  --0.525 &    0.847 &    2.989 &  --0.374 &    0.792 &    3.149 &  --0.517 &    0.842 \\
2.726 &  --0.185 &    0.941  &    2.582 &  --0.025 &    0.891 &    2.656 &  --0.033 &    0.803 &    2.668 &  --0.179 &    0.842 \\
1.756 &    0.178 &    0.492  &    2.273 &    0.142 &    0.383 &    1.930 &    0.348 &    0.990 &    2.446 &    0.074 &    0.891 \\
2.136 &    0.184 &    0.879  &    1.852 &    0.299 &    0.714 &          &          &          &    1.766 &    0.490 &    0.883 \\
1.676 &    0.507 &    0.865  &          &          &          &          &          &          &          &          &          \\
\end{tabular*}
\end{table*}

\begin{table*}
\addtocounter{table}{-1}
\caption{  -- \textit{continued}}
\begin{tabular*}{\linewidth}{c@{\hspace{1.em}}c@{\hspace{1.4em}}c@{\hspace{1.6em}}@{\hspace{3em}}c@{\hspace{1.em}}c@{\hspace{1.4em}}c@{\hspace{1.6em}}@{\hspace{3em}}c@{\hspace{1.em}}c@{\hspace{1.4em}}c@{\hspace{1.6em}}@{\hspace{3em}}c@{\hspace{1.em}}c@{\hspace{1.4em}}c@{\hspace{1.6em}}}
\toprule
$\log I_i$ & $\log\sigma_i$ & $q_i$ & $\log I_i$ & $\log\sigma_i$ & $q_i$ & $\log I_i$ & $\log\sigma_i$ & $q_i$ & $\log I_i$ & $\log\sigma_i$ & $q_i$\\
$[L_{\odot\,r}\,\mathrm{pc}^{-2}]$ & [arcsec] & & $[L_{\odot\,r}\,\mathrm{pc}^{-2}]$ & [arcsec] & & $[L_{\odot\,r}\,\mathrm{pc}^{-2}]$ & [arcsec] & & $[L_{\odot\,r}\,\mathrm{pc}^{-2}]$ & [arcsec]&\\[1pt]
\midrule
\multicolumn{12}{c}{} \\  
\multicolumn{3}{c@{\hspace{3em}}}{J1029+0420}& \multicolumn{3}{c@{\hspace{3em}}}{J1032+5322}& \multicolumn{3}{c@{\hspace{3em}}}{J1103+5322} & \multicolumn{3}{c@{\hspace{3em}}}{J1106+5228} \\
\\
  4.279 &  --1.532 &    0.736 &   4.256 &  --1.532 &    0.827  &    3.779 &  --1.532 &    0.744 &    4.608 &  --1.532 &    0.644 \\    
  3.788 &  --0.998 &    0.758 &   3.805 &  --0.991 &    0.842  &    3.501 &  --1.012 &    0.812 &    3.928 &  --0.975 &    0.743 \\    
  3.493 &  --0.715 &    0.714 &   3.450 &  --0.660 &    0.812  &    3.369 &  --0.622 &    0.677 &    3.734 &  --0.961 &    0.545 \\    
  3.125 &  --0.454 &    0.821 &   2.745 &  --0.298 &    0.813  &    3.057 &  --0.083 &    0.347 &    3.710 &  --0.741 &    0.743 \\    
  2.962 &  --0.195 &    0.396 &   2.789 &  --0.038 &    0.297  &    2.228 &    0.020 &    0.842 &    2.948 &  --0.580 &    0.545 \\    
  2.065 &    0.075 &    0.891 &   2.327 &    0.188 &    0.348  &    2.441 &    0.234 &    0.347 &    3.459 &  --0.482 &    0.743 \\    
  2.622 &    0.094 &    0.396 &   1.288 &    0.390 &    0.842  &    1.918 &    0.312 &    0.574 &    2.785 &  --0.240 &    0.545 \\    
  2.198 &    0.299 &    0.504 &   1.577 &    0.390 &    0.297  &    1.369 &    0.555 &    0.447 &    2.941 &  --0.020 &    0.626 \\    
  1.411 &    0.517 &    0.513 &         &          &           &    1.314 &    0.555 &    0.842 &    1.996 &    0.279 &    0.743 \\
  1.502 &    0.517 &    0.891 &         &          &           &          &          &          &    2.211 &    0.350 &    0.545 \\
        &          &          &         &          &           &          &          &          &    1.787 &    0.678 &    0.723 \\ 
\\        
\multicolumn{3}{c@{\hspace{3em}}}{J1112+0826}&\multicolumn{3}{c@{\hspace{3em}}}{J1134+6027}& \multicolumn{3}{c@{\hspace{3em}}}{J1142+1001}& \multicolumn{3}{c@{\hspace{3em}}}{J1143--0144}  \\
\\
  3.734 &  --1.240 &    0.792 &   4.047 &  --1.532 &    0.743  &   3.710  &  --1.532 &   0.990  &    3.350 &  --0.792 &    0.832 \\    
  3.535 &  --0.763 &    0.743 &   3.784 &  --0.920 &    0.743  &   3.692  &  --1.059 &   0.990  &    3.507 &  --0.515 &    0.743 \\    
  3.070 &  --0.630 &    0.743 &   3.346 &  --0.597 &    0.743  &   3.403  &  --0.771 &   0.990  &    3.213 &  --0.228 &    0.743 \\    
  3.099 &  --0.368 &    0.743 &   3.045 &  --0.348 &    0.820  &   2.928  &  --0.495 &   0.990  &    2.732 &    0.156 &    0.753 \\    
  2.689 &  --0.072 &    0.755 &   2.387 &    0.012 &    0.743  &   2.690  &  --0.221 &   0.990  &    2.095 &    0.436 &    0.891 \\    
  2.142 &    0.325 &    0.792 &   2.218 &    0.226 &    0.879  &   2.105  &    0.176 &   0.771  &    1.507 &    0.919 &    0.813 \\    
        &          &          &   1.522 &    0.674 &    0.891  &   1.812  &    0.527 &   0.743  &          &          &          \\    
\\
 \multicolumn{3}{c@{\hspace{3em}}}{J1153+4612}&
\multicolumn{3}{c@{\hspace{3em}}}{J1204+0358}&\multicolumn{3}{c@{\hspace{3em}}}{J1205+4910}&
\multicolumn{3}{c@{\hspace{3em}}}{J1213+6708} \\
\\
  4.166 &  --1.532 &    0.842 &   4.089 &  --1.532 &    0.891 &     3.819 &  --1.532 &    0.842 &   4.345  &  --1.532 &    0.857  \\ 
  3.857 &  --1.027 &    0.842 &   3.766 &  --0.950 &    0.891 &     3.386 &  --1.139 &    0.842 &   4.110  &  --1.052 &    0.822  \\ 
  3.353 &  --0.742 &    0.842 &   3.275 &  --0.614 &    0.891 &     3.678 &  --0.921 &    0.842 &   3.635  &  --0.729 &    0.891  \\ 
  2.869 &  --0.527 &    0.842 &   2.778 &  --0.347 &    0.936 &     3.134 &  --0.628 &    0.842 &   3.170  &  --0.424 &    0.880  \\ 
  2.637 &  --0.381 &    0.990 &   2.533 &  --0.038 &    0.990 &     2.923 &  --0.433 &    0.693 &   2.744  &  --0.007 &    0.779  \\ 
  2.588 &  --0.112 &    0.990 &   1.992 &    0.345 &    0.990 &     2.732 &  --0.157 &    0.842 &   2.147  &    0.301 &    0.799  \\ 
  1.852 &    0.391 &    0.842 &         &          &          &     2.371 &    0.195 &    0.693 &   1.702  &    0.417 &    0.693  \\ 
        &          &          &         &          &          &     1.732 &    0.589 &    0.693 &   1.647  &    0.805 &    0.693  \\ 
        &          &          &         &          &          &     0.827 &    0.589 &    0.842 &          &          &           \\ 
\\
 \multicolumn{3}{c@{\hspace{3em}}}{J1218+0830} & \multicolumn{3}{c@{\hspace{3em}}}{J1250+0523}& \multicolumn{3}{c@{\hspace{3em}}}{J1402+6321}&\multicolumn{3}{c@{\hspace{3em}}}{J1403+0006} \\
\\
  3.347 &  --1.532 &    0.792 &   4.417 &  --1.532 &    0.990 &    3.875  &  --1.532 &    0.792 &    4.177 &  --1.532 &    0.891 \\  
  3.260 &  --1.031 &    0.792 &   3.871 &  --0.994 &    0.990 &    3.316  &  --1.058 &    0.842 &    3.638 &  --1.054 &    0.891 \\  
  3.383 &  --0.805 &    0.792 &   3.400 &  --0.591 &    0.990 &    3.537  &  --0.858 &    0.743 &    3.283 &  --0.680 &    0.891 \\  
  3.267 &  --0.542 &    0.693 &   2.818 &  --0.157 &    0.990 &    3.267  &  --0.603 &    0.743 &    1.979 &  --0.250 &    0.693 \\  
  3.024 &  --0.317 &    0.728 &   2.416 &    0.154 &    0.990 &    3.013  &  --0.368 &    0.743 &    2.899 &  --0.197 &    0.891 \\  
  2.668 &  --0.075 &    0.693 &   1.675 &    0.503 &    0.772 &    2.707  &  --0.059 &    0.796 &    2.197 &    0.174 &    0.709 \\  
  2.453 &    0.175 &    0.792 &         &          &          &    2.193  &    0.263 &    0.766 &    1.790 &    0.384 &    0.693 \\  
  1.807 &    0.475 &    0.720 &         &          &          &    1.751  &    0.619 &    0.842 &          &          &          \\  
  1.530 &    0.801 &    0.707 &         &          &          &           &          &          &          &          &          \\  
\\
 \multicolumn{3}{c@{\hspace{3em}}}{J1416+5136}& \multicolumn{3}{c@{\hspace{3em}}}{J1420+6019} & \multicolumn{3}{c@{\hspace{3em}}}{J1430+4105}& \multicolumn{3}{c@{\hspace{3em}}}{J1432+6317} \\ 
\\ 
3.661 &  --1.531 &    0.973  &    4.194 &  --1.532 &    0.743 &    3.902 &  --1.532 &    0.936 &    3.914 &  --1.532 &    0.976 \\
3.656 &  --1.039 &    0.990  &    3.936 &  --0.983 &    0.743 &    3.709 &  --0.998 &    0.941 &    3.634 &  --0.994 &    0.974 \\
3.296 &  --0.732 &    0.956  &    3.486 &  --0.671 &    0.743 &    3.404 &  --0.595 &    0.932 &    3.350 &  --0.678 &    0.978 \\
3.111 &  --0.352 &    0.794  &    3.384 &  --0.501 &    0.396 &    2.820 &    0.071 &    0.941 &    2.912 &  --0.358 &    0.907 \\
2.389 &    0.070 &    0.743  &    3.471 &  --0.219 &    0.396 &    1.707 &    0.652 &    0.594 &    2.732 &  --0.003 &    0.953 \\
1.664 &    0.427 &    0.743  &    3.024 &  --0.050 &    0.743 &          &          &          &    2.182 &    0.503 &    0.990 \\
1.090 &    0.427 &    0.990  &    2.435 &    0.172 &    0.396 &          &          &          &    0.158 &    0.811 &    0.396 \\
      &          &           &    2.176 &    0.299 &    0.743 &          &          &          &    1.435 &    0.811 &    0.990 \\
      &          &           &    2.031 &    0.452 &    0.436 &          &          &          &                                \\
      &          &           &    1.851 &    0.683 &    0.716 &          &          &          &                                \\         
\end{tabular*}
\end{table*}
        
\begin{table*}
\addtocounter{table}{-1}
\caption{  -- \textit{continued}}
\begin{tabular*}{\linewidth}{c@{\hspace{1.em}}c@{\hspace{1.4em}}c@{\hspace{1.6em}}@{\hspace{3em}}c@{\hspace{1.em}}c@{
\hspace{1.4em}}c@{\hspace{1.6em}}@{\hspace{3em}}c@{\hspace{1.em}}c@{\hspace{1.4em}}c@{\hspace{1.6em}}@{\hspace{3em}}c@{
\hspace{1.em}}c@{\hspace{1.4em}}c@{\hspace{1.6em}}}
\toprule
$\log I_i$ & $\log\sigma_i$ & $q_i$ & $\log I_i$ & $\log\sigma_i$ & $q_i$ & $\log I_i$ & $\log\sigma_i$ & $q_i$ & $\log I_i$ & $\log\sigma_i$ & $q_i$\\
$[L_{\odot\,r}\,\mathrm{pc}^{-2}]$ & [arcsec] & & $[L_{\odot\,r}\,\mathrm{pc}^{-2}]$ & [arcsec] & & $[L_{\odot\,r}\,\mathrm{pc}^{-2}]$ & [arcsec] & & $[L_{\odot\,r}\,\mathrm{pc}^{-2}]$ & [arcsec]&\\[1pt]
\midrule
\multicolumn{12}{c}{} \\
\multicolumn{3}{c@{\hspace{3em}}}{J1436--0000}& \multicolumn{3}{c@{\hspace{3em}}}{J1443+0304}& \multicolumn{3}{c@{\hspace{3em}}}{J1451--0239} & \multicolumn{3}{c@{\hspace{3em}}}{J1525+3327} \\
\\
  3.317 &  --1.532 &    0.792 &   4.369 &  --1.532 &    0.792  &    4.236 &  --1.532 &    0.952 &    3.969 &  --1.532 &    0.770 \\  
  3.227 &  --1.094 &    0.812 &   3.773 &  --1.045 &    0.990  &    3.756 &  --1.024 &    0.984 &    3.635 &  --1.030 &    0.792 \\  
  3.502 &  --0.783 &    0.772 &   3.462 &  --1.037 &    0.594  &    3.378 &  --0.645 &    0.921 &    3.279 &  --0.703 &    0.747 \\  
  3.125 &  --0.446 &    0.772 &   3.298 &  --0.774 &    0.594  &    2.936 &  --0.349 &    0.951 &    2.854 &  --0.469 &    0.594 \\  
  2.136 &  --0.054 &    0.772 &   3.194 &  --0.556 &    0.638  &    2.613 &  --0.074 &    0.990 &    2.801 &  --0.186 &    0.594 \\  
  2.434 &  --0.028 &    0.812 &   2.738 &  --0.299 &    0.594  &    1.716 &    0.186 &    0.297 &    2.559 &    0.130 &    0.671 \\  
  1.991 &    0.458 &    0.772 &   2.064 &  --0.039 &    0.990  &    2.202 &    0.213 &    0.990 &    1.944 &    0.574 &    0.606 \\  
        &          &          &   2.453 &  --0.022 &    0.594  &    1.690 &    0.621 &    0.990 &          &          &          \\  
        &          &          &   1.748 &    0.449 &    0.594  &          &          &          &          &          &          \\
        &          &          &   1.009 &    0.449 &    0.990  &          &          &          &          &          &          \\
\\
\multicolumn{3}{c@{\hspace{3em}}}{J1531--0105} &\multicolumn{3}{c@{\hspace{3em}}}{J1538+5817}& \multicolumn{3}{c@{\hspace{3em}}}{J1621+3931}& \multicolumn{3}{c@{\hspace{3em}}}{J1627--0053} \\
\\
  3.782 &  --1.532 &    0.718 &   4.324 &  --1.532 &    0.862 &     3.815 &  --1.443 &   0.767  &    3.864 &  --1.532 &    0.817 \\    
  3.753 &  --0.949 &    0.743 &   3.785 &  --0.991 &    0.862 &     3.546 &  --1.101 &   0.792  &    3.417 &  --1.144 &    0.792 \\    
  3.444 &  --0.699 &    0.693 &   3.393 &  --0.620 &    0.862 &     3.537 &  --0.910 &   0.743  &    3.627 &  --0.937 &    0.843 \\    
  3.216 &  --0.480 &    0.693 &   2.765 &  --0.212 &    0.822 &     3.339 &  --0.591 &   0.743  &    3.381 &  --0.572 &    0.827 \\    
  2.997 &  --0.280 &    0.693 &   2.432 &    0.070 &    0.852 &     2.884 &  --0.364 &   0.743  &    2.724 &  --0.085 &    0.792 \\    
  2.605 &  --0.033 &    0.693 &   1.751 &    0.475 &    0.853 &     2.712 &    0.003 &   0.792  &    2.172 &    0.125 &    0.941 \\    
  2.473 &    0.201 &    0.743 &         &          &          &     1.993 &    0.519 &   0.755  &    1.730 &    0.517 &    0.926 \\    
  1.836 &    0.709 &    0.693 &         &          &          &           &          &          &          &          &          \\
\\  
\multicolumn{3}{c@{\hspace{3em}}}{J1630+4520}& \multicolumn{3}{c@{\hspace{3em}}}{J1636+4707} &\multicolumn{3}{c@{\hspace{3em}}}{J2238--0754}& \multicolumn{3}{c@{\hspace{3em}}}{J2300+0022} \\
\\
  3.925 &  --1.532 &    0.831 &    3.922 &  --1.532 &    0.896 &   3.998 &  --1.532 &    0.733 &   2.876 &  --1.364 &    0.773  \\   
  3.796 &  --1.044 &    0.832 &    3.595 &  --0.991 &    0.941 &   3.668 &  --0.925 &    0.823 &   3.405 &  --0.982 &    0.843  \\   
  3.443 &  --0.648 &    0.831 &    3.334 &  --0.680 &    0.852 &   2.853 &  --0.668 &    0.644 &   3.280 &  --0.710 &    0.703  \\   
  2.917 &  --0.309 &    0.842 &    2.954 &  --0.422 &    0.941 &   3.234 &  --0.524 &    0.891 &   3.227 &  --0.513 &    0.833  \\   
  2.185 &  --0.164 &    0.684 &    2.365 &  --0.274 &    0.743 &   2.431 &  --0.279 &    0.644 &   2.342 &  --0.218 &    0.702  \\ 
  2.441 &    0.019 &    0.842 &    2.358 &  --0.129 &    0.941 &   2.713 &  --0.072 &    0.891 &   2.456 &  --0.067 &    0.990  \\   
  2.104 &    0.152 &    0.842 &    2.280 &    0.030 &    0.743 &   2.145 &    0.257 &    0.644 &   2.036 &    0.031 &    0.693  \\   
  1.865 &    0.478 &    0.842 &    1.965 &    0.447 &    0.743 &   1.797 &    0.627 &    0.644 &   1.567 &    0.424 &    0.990  \\   
        &          &          &          &          &          &         &          &          &   1.692 &    0.424 &    0.693  \\   
\\
\multicolumn{3}{c@{\hspace{3em}}}{J2303+1422} & \multicolumn{3}{c@{\hspace{3em}}}{J2321--0939}& \multicolumn{3}{c@{\hspace{3em}}}{J2341+0000} \\
\\
3.356 &  --1.202 &    0.743 &    3.920 &  --1.532 &    0.842 &    3.973 &  --1.523 &    0.754 \\
3.555 &  --0.816 &    0.743 &    3.682 &  --0.950 &    0.842 &    3.427 &  --0.766 &    0.716 \\
3.276 &  --0.440 &    0.698 &    3.656 &  --0.632 &    0.842 &    3.365 &  --0.652 &    0.792 \\
2.804 &  --0.023 &    0.685 &    3.199 &  --0.289 &    0.743 &    2.432 &  --0.128 &    0.495 \\
2.328 &    0.309 &    0.644 &    2.920 &    0.030 &    0.800 &    2.458 &    0.036 &    0.792 \\
1.774 &    0.693 &    0.646 &    1.896 &    0.254 &    0.743 &    2.311 &    0.325 &    0.495 \\
      &          &          &    2.434 &    0.422 &    0.805 &    1.548 &    0.369 &    0.792 \\
      &          &          &    1.724 &    0.846 &    0.776 &    1.525 &    0.697 &    0.792 \\
\bottomrule
\end{tabular*}
\flushleft
Notes: Column 1: Logarithm of the Gaussian amplitude. Column 2: Logarithm of the Gaussian width. Column 3: Axial ratio of the Gaussian.
\end{table*}

\label{lastpage}

\end{document}